\documentclass[twocolumn,aps,prc,superscriptaddress]{revtex4-2}
\usepackage{color,amsmath,amssymb,epsfig,graphicx}
\usepackage{bm}% bold math
\usepackage[colorlinks=true,linkcolor=blue,filecolor=blue,urlcolor=blue,citecolor=blue]{hyperref}
\usepackage{mathptmx, courier, pifont}
\usepackage[scaled=0.92]{helvet}
\usepackage[T1]{fontenc}
\usepackage{textcomp}
\usepackage{multirow}
\usepackage{longtable}

\begin{document}

%\title{ E2 Transitions and Shape Invariants}
\title{ Microscopic investigation of $E2$ matrix elements in atomic nuclei }

\author{S.P. Rouoof} \email{sprouoofphysics27@gmail.com}
\affiliation{Department of  Physics, Islamic University of Science and Technology, Awantipora, 192 122, India}
\author{Nazira Nazir}
\affiliation{Department of Physics, University of Kashmir, Srinagar, 190 006, India}
\author{S.~Jehangir}
\affiliation{Department of Physics, Government Degree College for women, Pulwama, Jammu and Kashmir, 192 301, India}
\author{G.H.~Bhat}
\affiliation{Department of Higher Education (GDC Shopian), Jammu and Kashmir, 192 303, India}
\author{J.A. Sheikh} \email{sjaphysics@gmail.com}
\affiliation{Department of  Physics, Islamic University of Science and Technology, Awantipora, 192 122, India}
\affiliation{Department of Physics, University of Kashmir, Srinagar, 190 006, India}
\author{N. Rather}
\affiliation{Department of  Physics, Islamic University of Science and Technology, Awantipora, 192 122, India}
\author{S.~Frauendorf}
\affiliation{Department of Physics, University of Notre Dame, Notre Dame, Indiana 46556,  USA}

\date{\today}

\begin{abstract}

A systematic analysis of $E2$ matrix elements of $^{72}$Ge, $^{76}$Ge, $^{168}$Er, $^{186}$Os, $^{188}$Os, $^{190}$Os,
$^{192}$Os and  $^{194}$Pt nuclides is performed using the beyond mean-field approach of triaxial projected shell
model (TPSM). For these nuclei, large sets of $E2$ matrix elements have been deduced from the
multi-step Coulomb excitation experiments, and it is shown that TPSM approach provides a reasonable description
of the measured transitions. We have evaluated 1496 $E2$ matrix elements up to spin, $I=10$
for the eight nuclei studied, and tabulate them for
future experimental and theoretical comparisons. Further, shape invariant analysis
has been performed with the calculated $E2$ transitions using the Kumar-Cline sum rules. It is inferred from the
analysis that the resulting shape, after configuration mixing of the quasiparticle states, transforms
from $\gamma$-rigid to that of $\gamma$-soft for some nuclei, in conformity with the experimental data. 
   
\end{abstract}

%\pacs{21.60.Cs, 23.20.Lv, 23.20.-g, 27.70.+q}

\maketitle

\section{Introduction}

The Coulomb excitation (COULEX) mechanism provides an invaluable information on the $E2$ matrix elements of the
nuclear states \cite{D.Cline.Annu.Rev.Nucl.Sci}. In the early days of nuclear physics, it
was possible to populate only low-spin states
with light-ion beams. However, the availability of the heavy-ion beams coupled with the development
of the high-resolution gamma-ray detector arrays has led to a renaissance in the field of COULEX. Further,
very advanced computer codes, for instance, GOSIA \cite{GOSIA2012} have been developed for analysing
the vast amounts of data collected from
these advanced COULEX experiments. It is now possible to extract a complete set of $E2$ matrix elements for low-lying
states up to an excitation energy of 2 MeV from the COULEX data. 

The extraction of a large
set of $E2$ matrix elements for a given nucleus has made it possible to infer the shape of the nucleus in a model independent way
by calculating the quadrupole shape invariants  \cite{D.Cline.Annu.Rev.Nucl.Sci}.
It is known that electric quadrupole spherical tensor operators can be expressed as products of rotationally
invariant quantities, which  can then be related to the ellipsoidal charge distribution of the nucleus in the intrinsic
frame of reference \cite{KM72}. The product of two- and three operators can be written in terms of
ellipsoidal shape parameters, similar to $\beta$ and $\gamma$ parameters of Bohr-Mottelson
collective model \cite{BMII}. The expectation values of the products of quadrupole operators can be completely expressed
in terms of $E2$ matrix elements of intermediate states involved in the summation. This method of evaluating
the rotational invariant quantities from measured $E2$ matrix elements provides a direct measure of the shape
of the equivalent ellipsoid in the intrinsic frame. In Bohr-Mottelson collective model, rotational $D$-functions
are employed to transform the quantities to the laboratory frame \cite{BMII}. 

Detailed  $E2$ matrix elements have been deduced from COULEX experiments in several regions of the mass table
\cite{Kotlinski1990,Ayangeakaa2016,D.Cline.Annu.Rev.Nucl.Sci,Wu1996,104Ru,Ayangeakaa2019}. In the mass
$A \sim 160$ region $^{168}$Er is probably one of the best studied systems with about 50 $E2$ matrix elements known for the
yrast- and the $\gamma$-bands \cite{Kotlinski1990}. The COULEX experiments on shape transitional Pt-Os region with $A \sim 190$
extracted $E2$ matrix elements  for several of these isotopes up to $I=12$ \cite{Wu1996,D.Cline.Annu.Rev.Nucl.Sci}. In the medium mass
region, $ A\sim 100$, extended sets of
$E2$ and $M1$ matrix elements in %$^{110}$Pd,
$^{104}$Ru, $^{72}$Ge and $^{76}$Ge have been measured by means of COULEX \cite{Ayangeakaa2016,Ayangeakaa2019,Ayangeakaa2023,104Ru}.

The evaluated $E2$ matrix elements have been compared with the two extreme limits of the Bohr-Mottelson collective
Hamiltonian, the rigid asymmetric rotor \cite{Davydov1958} and the  $\gamma$-independent potential
energy \cite{Wilets1956}. Further, the interacting boson model (IBM), with the Hamiltonian fitted
to the energy spectra, has
also been employed to investigate the measured $E2$ matrix elements \cite{AIannurevIBM,IBM,Otsuka1987}.
It has been shown that these three purely phenomenological approaches reproduce only
some aspects of the known COULEX data \cite{Ayangeakaa2016,Kotlinski1990,D.Cline.Annu.Rev.Nucl.Sci,Ayangeakaa2023,Wu1996,Allmond2008,Ayangeakaa2019,104Ru,Warner1981,HARTER1987295}. 

The availability of the extended sets of $E2$ matrix elements represent a benchmark for assessing the validity and scope
of the microscopic approaches to describe the dynamics of the 
collective quadrupole degrees of freedom. Several nuclear structure models with some microscopic features have been
developed and here we only mention  
the microscopic Bohr Hamiltonian derived from quadrupole-quadrupole interaction, what is referred to as the quadrupole
constrained Bohr Hamiltonian (QCBH) \cite{KZ99}.  
 %This approach has been recently
%extended to include neutron and proton pairing vibrational modes as dynamical variables, and it has
%been shown to reproduce some collective excited states, which was one of the major
%drawbacks of the earlier version \cite{KZ99}. This advanced model referred to as quadrupole
%constrained Bohr Hamitonian (QCBH) 
This model has been used to describe the COULEX  $E2$ matrix elements for several  nuclides with varying success
\cite{Kotlinski1990,Ayangeakaa2016,Wu1996,104Ru,Ayangeakaa2019}.
% Further, mapping procedures
%have been utilised to determine the parameters of the Bohr-Mottelson
%Hamitonian from effective mean-field theories \cite{KZ99}. The mapped Hamiltonian is then used to study
%low-spin properties of nuclei, including the $E2$ matrix elements.

The purpose of the present work is to employ the large sets of E2
matrix elements for eight nuclides, available from the COULEX data, as benchmarks to 
appraise the performance of the beyond mean-field  microscopic approach of the triaxial projected shell model (TPSM), which extends
our earlier detailed study
on $^{104}$Ru along these lines \cite{nazira}. In contrast to microscopic versions of the Bohr Hamiltonian,
the TPSM approach does not introduce an explicit {\it ansatz} for the collective quadrupole degrees of freedom, and avoids
invoking the adiabatic approximation.
The model rather represents a shell model based approach with angular momentum projected deformed configurations
as the basis states. The deformed states contain essential correlations needed to describe the deformed systems,
and the collective features emerge directly from the
microscopic degrees of freedom. As this approach includes multi-quasiparticle configurations, it is capable of describing 
the properties of nuclei up to quite high-spin. In several 
studies, it has been shown that TPSM approach provides a remarkable description of the single-particle
and collective excitations in deformed and transitional systems \cite{SH99,JS16,JS21}. The phenomena of chiral symmetry and wobbling
motion have been investigated
in detail in the TPSM approach \cite{JS16}.
Recently, a systematic study of the $\gamma$-bands observed in atomic nuclei have been undertaken, and it has
been shown that staggering phase of these bands can change from the pattern of $\gamma$-rigid to
that of $\gamma$-soft with the inclusion of the quasiparticle excitations \cite{JS21,Rouoof2024}.

In our more recent work \cite{nazira}, we calculated
  a complete set of $E2$ matrix elements,
  the centroids and dispersions of the shape invariants   for $^{104}$Ru  were investigated. Our work demonstrated
  that, although the triaxial mean-field in the TPSM approach is $\gamma$-rigid, the beyond-mean-field procedure
  of admixing  angular-momentum projected multi-quasiparticle states to the projected vacuum configuration
  alters the characteristics of $^{104}$Ru from $\gamma$-rigid
  to $\gamma$-soft as deduced from the measured $E2$ matrix elements. In the present manuscript, the analysis carried out
  for $^{104}$Ru is broadened for the eight nuclei of $^{72}$Ge, $^{76}$Ge, $^{168}$Er, $^{186}$Os, $^{188}$Os, $^{190}$Os,
  $^{192}$Os and  $^{194}$Pt. We have chosen these systems as extensive $E2$ matrix elements have been measured
  for these systems \cite{Kotlinski1990,Ayangeakaa2016,Wu1996,Ayangeakaa2019,Ayangeakaa2023}. The remaining
  manuscript is organized in the following manner. The TPSM
  approach and shape invariant sum rule analysis are briefly discussed for completeness in sections \ref{Sect.02} and \ref{SISRA} respectively. In section \ref{RAD}, the results of $E2$ matrix elements and shape
  invariant quantities are
  presented and compared with the known data \cite{Kotlinski1990,Ayangeakaa2016,Wu1996}. In section \ref{SAC}, the present work
  is summarized and concluded.

\section{
Triaxial Projected Shell Model Approach}
\label{Sect.02}

TPSM approach has its roots in the pairing plus quadrupole model (PPQM)
of Kumar and Baranger \cite{Kumar1968}. PPQM is the microscopic formulation of 
the well-known collective model of Bohr and Mottelson (BM) with quadrupole and pairing
degrees of freedom as the basic building blocks \cite{BMII}. There have been several studies
demonstrating that low-lying properties of atomic nuclei can be described reasonably well using
the PPQM approach \cite{Xu2009}. In the seventies and eighties, high-spin properties of rotating nuclei were systematically
investigated using the PPQM Hamiltonian by applying the cranking \cite{BANERJEE1973} and angular-momentum
projection techniques \cite{KY95}. However, most of these studies were restricted to investigate the properties
of rotating nuclei close to the yrast configuration \cite{HARA1991445}. In the deformed $A \sim 160$ region,
more than twenty side-bands have been identified in some nuclei and in order to elucidate the properties of these
rich band structures, it is imperative to include quasiparticle excited configurations in the model. The semiclassical
cranked shell model (CSM) \cite{Ben79aw} introduced the concept of quasiparticles  in a rotating potential, which allowed the
authors to classify the complex band structures as independent quasiparticle configurations of these building blocks. However,
in this approach, the quasiparticle states don't have well defined angular-momentum and predictions of transition probabilities become less reliable, in particular for
interband transitions.

In an attempt to investigate the rich band structures observed in deformed nuclei, projected shell model
(PSM) approach has been developed \cite{KY95}. In this approach, the multi-quasiparticle configurations are
generated from deformed Nilsson potential and the Bardeen-Cooper-Schriffer (BCS) approximation. The Nilsson potential
is directly employed as the quadrupole mean-field instead of solving the Hartree-Fock (HF) equation
with quadrupole-quadrupole interaction \cite{KY95}. This approach has the advantage that the Nilsson parameters have
been fitted to a large set of the experimental data, and is known to provide an accurate description of the
ground state properties of deformed nuclei \cite{Ni69,Moller1995}. The deformed Nilsson basis is then employed to construct the
multi-quasiparticle states with the use of the BCS formalism. In the second
step, the quasiparticle configurations are  projected onto good angular-momentum states using the angular-momentum projection
operator \cite{KY95}. In the final stage, the projected states are  used to diagonalize the shell model Hamiltonian
consisting of pairing (monopole and quadrupole) and quadrupole-quadrupole interaction terms. This approach is similar to the spherical shell model
(SSM) approach with the difference that deformed basis is used in comparison to the spherical basis
used in SSM. The deformed basis constitute the optimum basis to investigate a deformed system, and the TPSM analysis needs
only about 60 quasiparticle states  to describe a deformed nucleus up to quite high-spin.
%In comparison to the
%simpler CSM, the TPSM avoids problems due to non-conservation of angular momentum at the price of increasingly complex
%configurations of non-rotating quasiparticles.

In the original version of the projected shell model approach \cite{HS79,HS80,Iwasaki1982,HARA1984175,SUN1997245,rp01}, the deformed basis were restricted
to have axial symmetry and transitional nuclei
could not be studied using this approach. The model was generalized to include triaxial
basis states, which are obtained by solving the three dimensional Nilsson potential \cite{SH99}. 
This extended version of the model, what is referred to as
the TPSM approach, has been shown to describe the properties of transitional nuclei remarkably well \cite{SH99,JS16,JS21,Jeh2022} and
the reason is that the TPSM incorporates the collective features caused by deviations from axial shape.

In the present work, we have used the extended TPSM version \cite{SH99,JS16,JS21,Jeh2022} to investigate the detailed $E2$ matrix of even-even
isotopes.
The projected basis states
considered for studying the even-even system are the
vacuum, the two-proton, the two-neutron and the two-proton plus two-neutron
quasiparticle configurations, i.e.,
\begin{eqnarray}
\{ \hat P^I_{MK}\left|\Phi\right>, ~\hat P^I_{MK}~a^\dagger_{p_1}
a^\dagger_{p_2} \left|\Phi\right>, ~\hat P^I_{MK}~a^\dagger_{n_1}
a^\dagger_{n_2} \left|\Phi\right>,  \nonumber \\~\hat
P^I_{MK}~a^\dagger_{p_1} a^\dagger_{p_2} a^\dagger_{n_1}
a^\dagger_{n_2} \left|\Phi\right> \}, \label{basis}
\end{eqnarray}
where $\left|\Phi\right>$ in (\ref{basis}) represents the triaxial quasiparticle 
vacuum state. In majority
of the nuclei, near-yrast spectroscopy up to $I=20$ is well described
using the above basis space. 

The quasiparticle states are the standard BCS ones for the single particle states 
of the Nilsson Hamiltonian \cite{Nilsson1969},
\begin{equation}
\hat H_N = \hat H_0 - {2 \over 3}\hbar\omega\left\{\epsilon\hat Q_0
+\epsilon'{{\hat Q_{+2}+\hat Q_{-2}}\over\sqrt{2}}\right\}.
\label{nilsson}
\end{equation}
The term $\hat H_0$ contains the spherical single-particle energies,
the parameters of which are fitted to a broad
range of nuclear properties \cite{KY95}. The axial deformation parameter $\varepsilon$ and
triaxiality parameter $\varepsilon'$ (or $\gamma=\arctan(\varepsilon'/\varepsilon$) are
fixed input parameters of the model. They are adjusted to reproduce the reduced transition  probability
$B(E2, 2^+_1\rightarrow 0^+_1 )$ and the energy
$E(2^+_2)$ of the $\gamma$-band head. Table \ref{tab:parameters} provides these parameters for the nuclides
studied in the present work.

The  basis (\ref{basis}) is
employed to diagonalize the TPSM Hamiltonian, which
 consists of pairing and quadrupole-quadrupole interaction
terms, i.e.,
\begin{equation}
\hat H =  \hat H_0 -  {1 \over 2} \chi \sum_\mu \hat Q^\dagger_\mu
\hat Q^{}_\mu - G_M \hat P^\dagger \hat P   - G_Q \sum_\mu \hat
P^\dagger_\mu\hat P^{}_\mu . \label{hamham}
\end{equation}

The QQ-force strength, $\chi$, in Eq. (\ref{hamham}) is determined from
the quadrupole deformation $\epsilon$ as a result of the
self-consistency HFB condition  
\cite{KY95}:
\begin{equation}
\chi_{\tau\tau'} =
{{{2\over3}\epsilon\hbar\omega_\tau\hbar\omega_{\tau'}}\over
{\hbar\omega_n\left<\hat Q_0\right>_n+\hbar\omega_p\left<\hat
Q_0\right>_p}},\label{chi}
\end{equation}
where $\omega_\tau = \omega_0 a_\tau$, with $\hbar\omega_0=41.4678
A^{-{1\over 3}}$ MeV, and the isospin-dependence factor $a_\tau$ is
defined as
\begin{equation}
a_\tau = \left[ 1 \pm {{N-Z}\over A}\right]^{1\over 3},\nonumber
\end{equation}
with $+$ $(-)$ for $\tau =$ neutron (proton). The harmonic
oscillation parameter is given by $b^2_\tau=b^2_0/a_\tau$ with
$b^2_0=\hbar/{(m\omega_0)}=A^{1\over 3}$ fm$^2$. Note
that the strengths in the TPSM are fixed as in the original PSM with axial symmetry \cite{KY95}.
Possible corrections from the non-axial component of the quadrupole field 
have been neglected in the present work.

%=====================table parameters one===================================================================
\begin{table*}[htp!]
%\LTcapwidth=0.4\textwidth       
\caption{\label{tab:parameters}Axial and triaxial quadrupole deformation parameters
  $\epsilon$ and $\epsilon'$  employed in the TPSM calculation. Axial deformations are taken from our earlier
  works \cite{JS21, GH14, GH08, bh15, GS12} and \cite{raman},
and nonaxial deformations are chosen in such a way that heads
of the $\gamma$-bands are reproduced.}
%% \resizebox{1.0\textwidth}{!}
%% {
\begin{tabular}{p{2.0cm}p{2.0cm}p{2.0cm}p{2.0cm}p{2.0cm}p{2.0cm}p{2.0cm}p{2.0cm}c}%{M{2cm}M{2cm}M{2cm}M{2cm}M{2cm}M{2cm}M{2cm}M{2cm}M{2cm}M{2cm}N}
  \hline\hline
%%===================================================================									
 Isotope  	&$^{72}$Ge	&$^{76}$Ge	&$^{168}$Er	&$^{186}$Os	&$^{188}$Os	&$^{190}$Os 	&$^{192}$Os	&$^{194}$Pt	\\
  \hline									
$\epsilon$	&0.230	&0.200	&0.321	&0.200	&0.183	&0.178	&0.164	&0.125	\\
$\epsilon'$	&0.160	&0.160	&0.130	&0.118	&0.088	&0.092	&0.085	&0.073	\\
%%===================================================================									

\hline\hline								

\end{tabular}
%}
\end{table*}
%%=======================================================================================

The monopole pairing strength $G_M$ (in MeV)
is of the standard form
\begin{eqnarray}
G_M = {{G_1 \mp G_2{{N-Z}\over A}}\over A}, 
 \label{pairing}
\end{eqnarray}
where the minus (plus) sign applies to neutrons (protons). 
In the present calculation, we choose  $G_1$ and $G_2$
such that the calculated gap parameters reproduce the experimental
mass  differences. The values $G_1$ and $G_2$ vary depending
on the mass region and shall be listed in the discussion
of the results. The above choice of $G_M$ is appropriate for the
single-particle space employed in the model, where three major
shells are used for each type of nucleon. For most of the nuclei, 
Eq.~(\ref{pairing}) has been employed. However, it has been found that
for protons, $G_M = {G_1 \over A} $ reproduces the pairing gaps for 
$A\sim 130$ region \cite{bh14a} slightly better and has been 
employed in this region. The quadrupole pairing strength $G_Q$ is
considered to be proportional to $G_M$ and the proportionality
constant being fixed as 0.18. These interaction strengths are
consistent with those used in our earlier studies
\cite{GH08,JG09,SB10,JG11,JG12,YK00,KY95,JY01,js01,rp01,Ch12,YS08}.

The projection formalism outlined above
can be transformed into a diagonalization problem following the
Hill-Wheeler prescription. The Hamiltonian in Eq. (\ref{hamham}) is diagonalized using the projected 
basis of Eq. (\ref{basis}). 
 The obtained wavefunction can be written as
\begin{equation}
\psi^{\sigma I}_{IM} = \sum_{K,\kappa}f^{\sigma}_{\kappa K}\hat P^{I}_{MK}| 
~ \Phi_{\kappa} \rangle
\label{17}
\end{equation}
Here, the index $^{``}\sigma^{"}$ labels the eigenstates for a given  angular momentum
and $^{``}\kappa^{"}$ the quasiparticle configurations of the basis states. In Eq. (\ref{17}),
$f^{\sigma}_{\kappa K}$ are the amplitudes of the basis states
$\kappa , K$.
These wavefunctions are used to calculate matrix element of a multipole operator
$\hat {\mathcal {O}}_L$ given by
\begin{align}
  \langle \psi^{\sigma_f}_{I_fM_f}|\hat {\mathcal {O}}_{LM}&|\psi^{\sigma_i}_{I_iM_i}\rangle\nonumber\\
  &= (I_iM_i,LM|I_fM_f)\langle \sigma_f , I_f || \hat {\mathcal {O}}_L|| \sigma_i , I_i\rangle, 
\label{E22}
\end{align}
where the reduced matrix element of an operator $\hat {\mathcal {O}}$ can be expressed as \cite{Sun94}
\begin{align}
%\begin{split}
\langle \sigma_f , I_f || \hat {\mathcal {O}}_{\lambda} &||\sigma_i , I_i\rangle 
\nonumber \\
& = \sum_{\kappa_i,K_i , \kappa_f, Kf} {f_{ \kappa_i K_i}^{\sigma_i I_i}}~~~{f_{ \kappa_f K_f}^{\sigma_f I_i}}
 \sum_{M_i , M_f , M} (-)^{I_f - M_f}  \nonumber \\&~~~~~~~\times
\left( \begin{array}{ccc}
 I_f & L & I_i \\
-M_f & M & M_i 
\end{array} \right) \nonumber \\
 &~~~~~~~\times  \langle \Phi_{\kappa_f} | {\hat{P}^{I_f}}_{K_f M_f} \hat {\mathcal {O}}_{LM}
\hat{P}^{I_i}_{K_i M_i} | \Phi_{\kappa_i} \rangle  \nonumber \\
&= 2 \sum_{\kappa_i , \kappa_f} {f_{ \kappa_i K_i}^{\sigma_i I_i}}~~~{f_{ \kappa_f K_f}^{\sigma_f I_f}}\sum_{M' , M''} (-)^{I_f - K_{\kappa_f}} {(2I_f + 1)}^{-1}\nonumber\\
&~~~~~~~\times  \left( \begin{array}{ccc}
 I_f & L & I_i \\
-K_{f} & M' & M'' 
\end{array} \right)~\int d\Omega \,D^{I_i}_{M''K_{i}}(\Omega)\nonumber\\
&~~~~~~~\times\langle \Phi_{\kappa_f}|\hat {\mathcal {O}}_{LM'}\hat R(\Omega)|\Phi_{\kappa_i}\rangle.
\label{beeee2}
%\end{split} 
\end{align}

\section{Invariant Sum Rule Analysis}
\label{SISRA}

The individual $E2$ matrix elements are strongly correlated by the  quadrupole collectivity, and the number of important collective
variables is 
substantially lower than the number of matrix elements \cite{D.Cline.Annu.Rev.Nucl.Sci}. The conventional method of comparing the TPSM results 
to the set of experimental $E2$ matrix elements demonstrates the capability of the model
to account for the data. 
%does not demonstrate the uniqueness or sensitivity of the data to the collective characteristics.
%Furthermore, these comparisons cannot determine if the differences between experimental and theoretical values are the result of a
%basic flaw in the model or merely insufficiently accurate collective model parameters have been utlilized. 
The projection of the
collective characteristics 
from both the data and the microscopic calculations provides a greater 
insight. 
%since it makes evident as to which collective
%parameters are determined by the data and how good the collective model description is. 
Following the work of Kumar and
Cline,  the experimental electromagnetic
matrix elements can be interpreted in terms of collective degrees of freedom for any state by using the expectation values of rotationally invariant products of multipole operators which relate properties in the principal axis frame to those in the
laboratory frame\cite{D.Cline.Annu.Rev.Nucl.Sci,KM72}. 

The spherical tensor nature of electromagnetic multipole operators makes it possible to construct zero-coupled products of such operators that are rotationally invariant.
This implies that in the intrinsic frame and the laboratory frame, these products are identical. For simplicity consider the case of the electric quadrupole operator $E2$. The intrinsic principal axis frame electric moments can be described in terms of two parameters ($Q,\delta$) by analogy with Bohr's shape parameters ($\beta,\gamma$) as 
%======================================================================
 \begin{eqnarray}
  &&E(2,0)=Q\textrm{cos}\delta\nonumber\\
  &&E(2,1)=E(2,-1)=0\\
  &&E(2,2)=E(2,-2)=\frac{1}{\sqrt{2}}.Q\textrm{sin}\delta\nonumber
 \end{eqnarray}
 %======================================================================
 where $Q$ and $\delta$ are the arbitrary parameters and $Q$ can be written in terms of intrinsic quadrupole moment for an axially symmetric shape as $Q=\sqrt{\frac{5}{16\pi}}Q_0$. In terms of the principal axis frame parameters, zero coupled products of $E2$ operators can be written as
%======================================================================
 \begin{eqnarray}
   &&[E2\times E2]^0=\frac{1}{\sqrt{5}}Q^2 \nonumber\\
   &&\left[[E2\times E2]^2\times E2 \right]^0=-\sqrt{\frac{2}{35}}Q^3\textrm{cos}3\delta. 
 \end{eqnarray}
 %======================================================================

 The symbol $[...\times ...]^L$ stands for vector coupling to angular momentum $L$. The coefficients in front of $Q^2$ and $Q^3$ are the corresponding products of the Wigner symbols. In order to have a correspondence with collective coordinates, these invariants are denoted here up to coefficients as $Q^2$ and $Q^3\textrm{cos}3\delta$. The expectation values of the $E2$ rotation invariants in the laboratory frame can be related to the reduced $E2$ matrix elements by making intermediate state expansions. The corresponding sum rules read:
%======================================================================
 \begin{align}
   \langle S|&[E2\times E2]^0|S\rangle\nonumber\\
  & =\frac{(-)^{2S}}{\sqrt{2S+1}}\sum_R\langle S||E2||R\rangle\langle R||E2||S\rangle\left\{ \begin{array}{ccc}
 2& 2 & 0 \\
S& S & R 
   \end{array}\right\}
 \end{align}  
\begin{align}   
   \langle S|[[&E2\times E2]^2 \times E2]^0|S\rangle\nonumber\\
   & =\sqrt{\frac{5}{2S+1}}\sum_{RT}\langle S||E2||R\rangle\langle R||E2||T\rangle\langle T||E2||S\rangle \nonumber\\
   &~~~~~~~~~~\times\left\{ \begin{array}{ccc}
 2& 2 & 0 \\
S& S & T 
\end{array}\right\}\left\{ \begin{array}{ccc}
2& 2 & 2 \\
T& S & R 
\end{array}\right\}(-)^{3S+T}
 \end{align}
 %======================================================================
where $S$ denotes state $S$ and at the same time the spin of state $S$ alone, $R$ and $T$ denotes the intermediate states and their spins, and $\{~~~~~~\}$ represents the $6j$ symbol. These intermediate states allows the rotational invariants built from $Q$ and $\delta$ to be expressed as the sums of the products of the reduced $E2$ matrix elements using the experimental data of these matrix elements with proper relative signs and magnitudes. This will directly determine the quantum distribution i.e., the centroids, dispersions, skewnesses, cross-correlation coefficients, etc., of $E2$ in a given state. The invariant $\langle S|Q^2|S \rangle$, i.e., expectation value of $Q^2$, denoted simply as $\langle Q^2\rangle$ is given by
 %======================================================================
 \begin{eqnarray}\label{Q2}
   \langle Q^2\rangle=\frac{1}{2S+1}\sum_R |\langle S||E2||R\rangle|^2
 \end{eqnarray}
 %======================================================================
 The invariant $\langle Q^2\rangle$ does not depend upon the sign of any matrix element, in contrast to invariant $\langle Q^3\textrm{cos}3\delta\rangle$ which is given by
%======================================================================
 \begin{align}\label{Q3}
 \langle Q^3&\textrm{cos}3\delta\rangle \nonumber\\ 
 & =\mp\sqrt{\frac{35}{2}}\frac{1}{2S+1}\sum_{RT}\langle S||E2||R\rangle\langle R||E2||T\rangle\langle T||E2||S\rangle \nonumber\\
   &~~~~~~~~~~~~~~~~~~~~~~~~~~~~~~~~~~~~~\times \left\{ \begin{array}{ccc}
 2& 2 & 2 \\
S& T& R 
\end{array}\right\}
 \end{align}
 %====================================================================== 
 where positive and negative signs correspond to integral and half-integral spin system respectively.

The dispersion of $Q^2$ is given by
%======================================================================
 \begin{eqnarray}
   \sigma(Q^2)=\sqrt{\langle Q^4\rangle-\langle Q^2\rangle^2}
 \end{eqnarray}
 %======================================================================
and the dispersion of the quadrupole asymmetry is given by
%======================================================================
 \begin{eqnarray}
   \sigma(\textrm{cos}3\delta)=\left[\frac{\langle \left(Q^3\textrm{cos}3\delta\right)^2\rangle}{\langle Q^6\rangle}-\left(\frac{\langle Q^3\textrm{cos}3\delta\rangle}{\langle Q^2\rangle^{3/2}}\right)^2\right]^{1/2}
 \end{eqnarray}
 %====================================================================== 

The calculation of the dispersions requires the evaluation of the 
higher order tensor products
$\langle Q^4\rangle$,$\langle Q^6\rangle$ and $\langle\left(Q^3\textrm{cos}3\delta\right)^2\rangle$. In analogy to Eqs. (\ref{Q2}) - (\ref{Q3}),
they are evaluated by multiple intermediate sums, the pertaining expression of which
are given in Ref.  \cite{D.Cline.Annu.Rev.Nucl.Sci}.  The sums  involve an increasing number of $E2$ matrix elements,
the limit being set  by the practical feasibility. 

%===========fig1========================================
\begin{figure*}[htb]
 \centerline{\includegraphics[trim=0cm 0cm 0cm
0cm,width=\textwidth,clip]{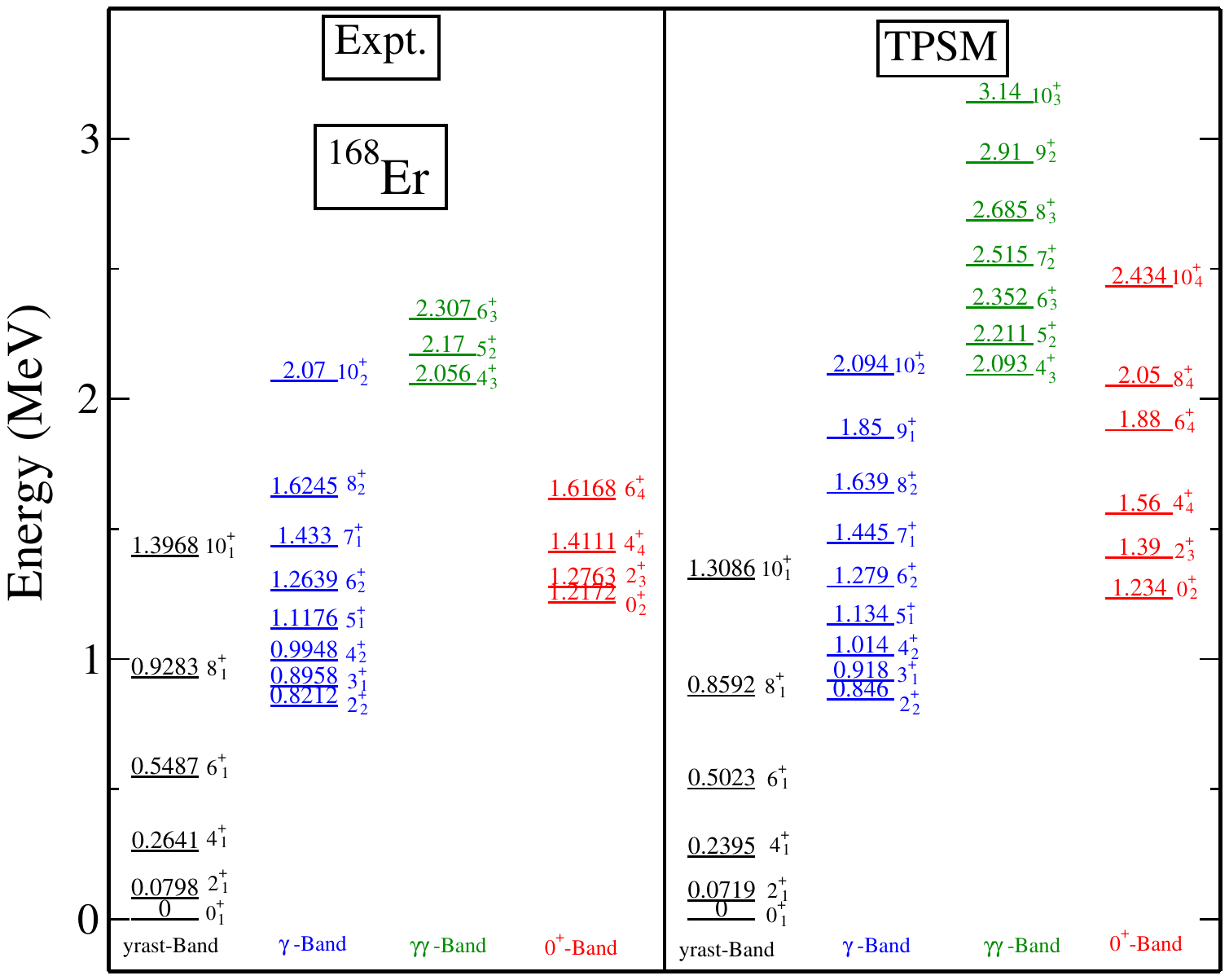}} \caption{(Color
online) TPSM and experimental energies of the lowest bands in  $^{168}$Er. The states are labelled by $I^+_n$.
These labels are used in the figures and tables to identify the states.  
  }
\label{E168Er}
\end{figure*}
%===================================================
 %===========fig2========================================
\begin{figure*}
    
 \centerline{\includegraphics[trim=0cm 0cm 0cm
0cm,width=0.8\linewidth,clip]{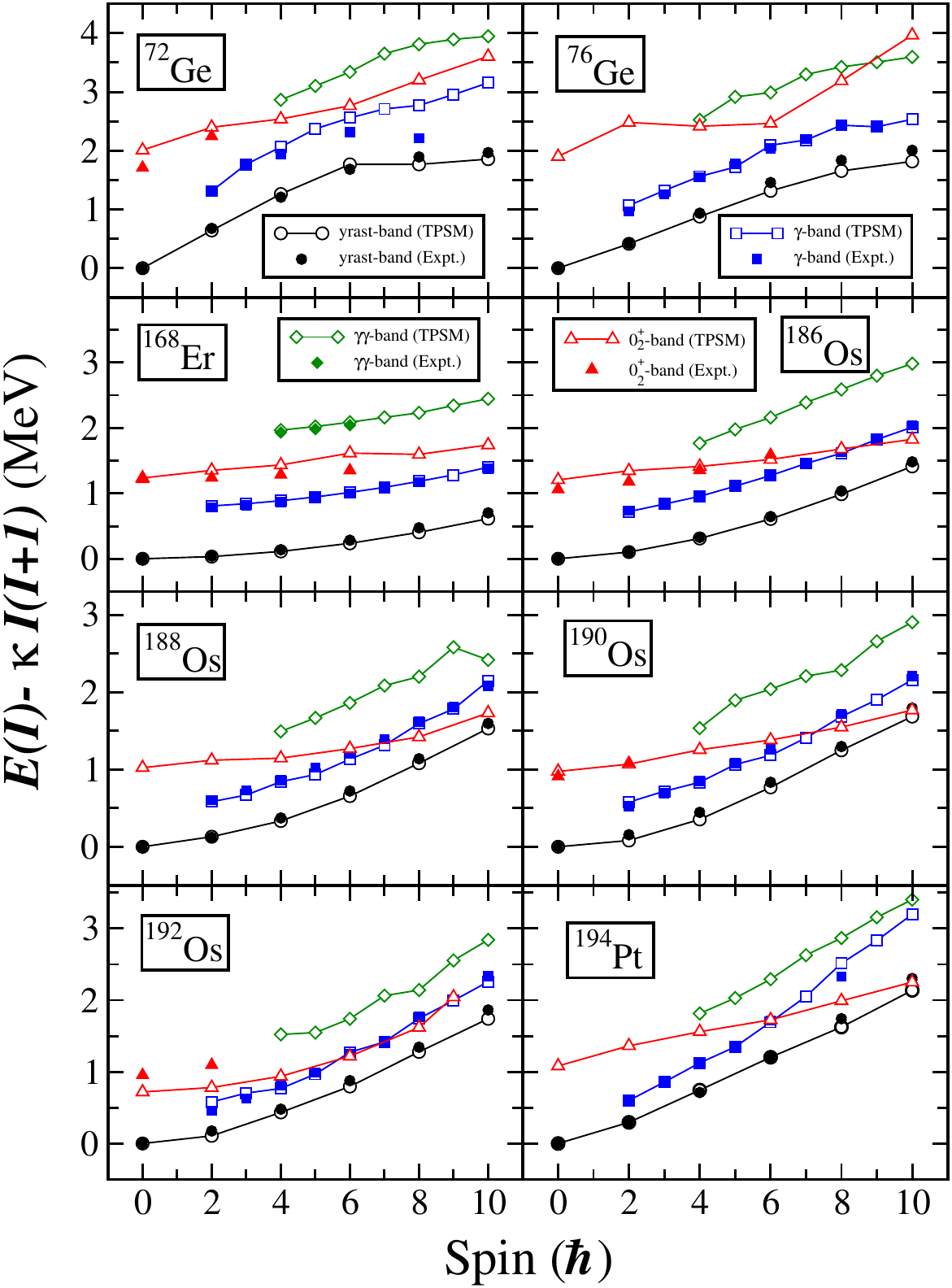}} \caption{(Color
online) TPSM and experimental energies of the yrast-, $\gamma$-, $\gamma\gamma$- and $0_2^+$-bands of $^{72,76}$Ge, $^{168}$Er,$^{186,188,190,192}$Os and
$^{194}$Pt isotopes. The scaling factor  $\kappa=32.32A^{-5/3}$.
  }
\label{E-all}
\end{figure*}
%===================================================
 
 %===========fig3========================================
\begin{figure*}
    
 \centerline{\includegraphics[trim=0cm 0cm 0cm
0cm,width=0.8\linewidth,clip]{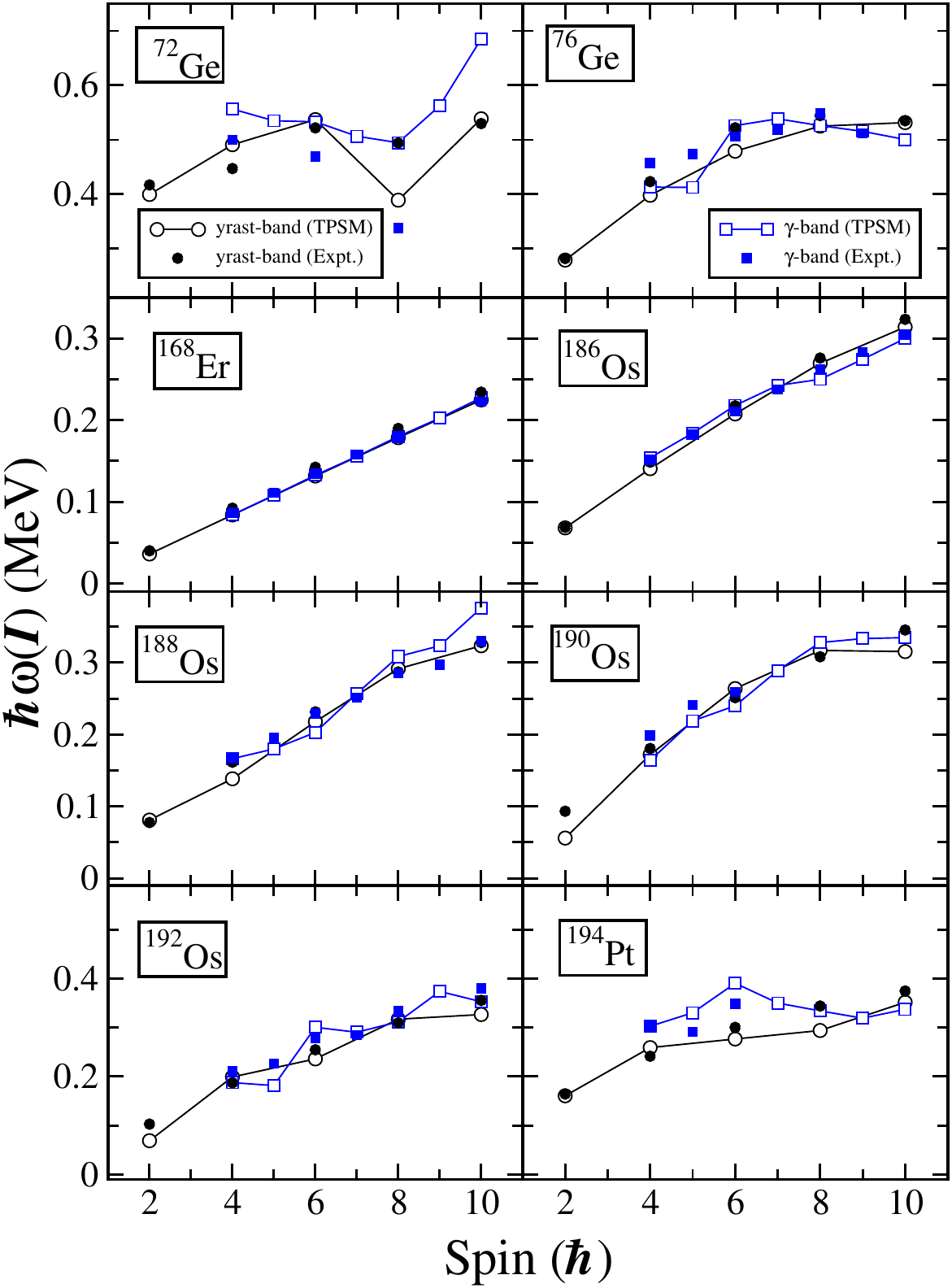}} \caption{(Color
online) TPSM and experimental $\hbar\omega (I)$ values of the yrast- and $\gamma$-bands of $^{72,76}$Ge, $^{168}$Er,$^{186,188,190,192}$Os and
$^{194}$Pt isotopes. Note that the experimental symbols cover the TPSM symbols.
  }
\label{omega-all}
\end{figure*}
%===================================================

%===========fig4========================================
\begin{figure*}[htb]
 \centerline{\includegraphics[trim=0cm 0cm 0cm
0cm,width=0.8\linewidth,clip]{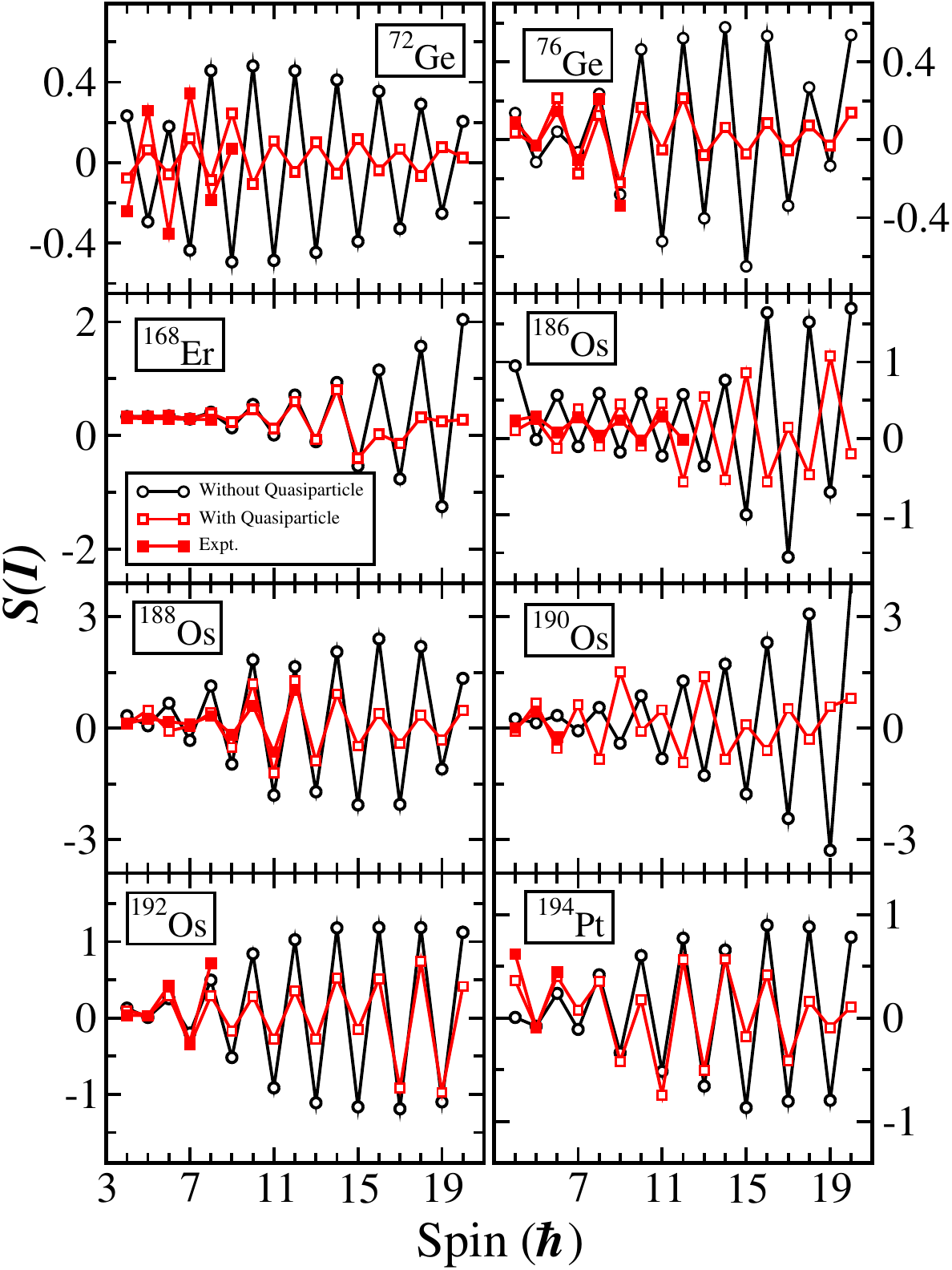}} \caption{(Color
online) Staggering parameters $S(I)$ of the $\gamma$-band in $^{72,76}$Ge, $^{168}$Er,$^{186,188,190,192}$Os and
$^{194}$Pt.
  }
\label{S-g}
\end{figure*}
%===================================================

\section{Results and Discussion}\label{RAD}

In this section, we shall present and discuss the TPSM results for the eight nuclides of
$^{72}$Ge, $^{76}$Ge, $^{168}$Er, $^{186}$Os, $^{188}$Os, $^{190}$Os,
$^{192}$Os and  $^{194}$Pt. We have evaluated both excitation energies and the $E2$ matrix elements, which are presented in the
following two subsections. In the third subsection, the analysis of the quadrupole shape invariants is discussed.

\subsection{Excitation Energies}
The primary objective of the present work is to investigate the $E2$ matrix elements. However, before
discussing these quantities, we shall first demonstrate that the TPSM approach reproduces the excitation energies
reasonably well for the eight nuclides studied.
As an illustrative example to demonstrate the robustness of the TPSM approach, Fig. \ref{E168Er} compares the experimental
level energies of the lowest four positive parity bands in $^{168}$Er with the TPSM
predictions. The results are depicted only for the low-lying states of the four bands that are populated in the COULEX experiments.
The band based on  the $2^+_2$ is the $\gamma$-band.  Its energy is adjusted to the experimental value by choosing the
appropriate value of the input triaxial parameter, $\varepsilon'$. 
The band based on  the $4^+_3$ level is called the $\gamma\gamma$-band throughout the paper. However, it needs to be added that it may be only a fragment of 
the collective double $\gamma$ vibration due to its coupling with
two-particle excitations. The TPSM approach accounts for the
fragmentation which will be discussed in detail in a separate work.

The excited $0^+$-band  with band head labeled as $0^+_2$ is higher in energy than the $\gamma$-band. 
For some nuclides studied  here, the excited $0^+$-band crosses the $\gamma$-band. However, irrespective of the position
of the states, we shall follow 
the labeling used in Fig. \ref{E168Er} throughout the manuscript. For instance, $6^+_4$ will always correspond to the
excited $0^+$-band.

For a detailed comparison between the experimental and TPSM energies, Fig. \ref{E-all} displays the excitation energies  for all eight studied nuclei with a common  reference
energy subtracted in order to expand the energy scale of the curves.
Evidently, the
TPSM reproduces the energies of the yrast- and $\gamma$-bands quite well. The energies of $\gamma\gamma$- and 
$0^+_2$-bands are also reasonably well described for the known states. However, 
the data is insufficient for these bands for
 a thorough comparison. A comprehensive comparison for more than 30 nuclides can be found in our previous
studies \cite{JS21,Rouoof2024}.

To probe the rotational properties of the systems, Fig. \ref{omega-all} compares the empirical rotational frequencies, defined as,
\begin{align}
    \hbar \omega(I)=\frac{E(I)-E(I-2)}{2}
\end{align}
with the corresponding TPSM values. Both the yrast- and the $\gamma$-band frequencies are
reproduced very well. The leveling  or down-bend of $\hbar \omega(I)$ at high spin 
is caused by the admixture of a pair of  rotationally aligned
$i_{13/2}$ quasineutrons in the case of the rare earth nuclei, 
and a pair of  aligned
$g_{9/2}$ quasineutrons in the case of the Germanium isotopes.

We have also evaluated the staggering parameter  of the $\gamma$-band,
defined as,
  \begin{eqnarray}\label{eq:staggering1}
S(I)= \frac{[E(I)-E(I-1)]-[E(I-1)-E(I-2)]}{E(2^{+}_1)}~~~~,
\end{eqnarray}
and displayed it in Fig. \ref{S-g}. The phase of $S(I)$ is used to classify 
 nuclei as "$\gamma$-rigid" when the odd $I$ levels are below the adjacent even $I$ levels or   "$\gamma$-soft"
 when the even $I$ levels are below
 the odd $I$ states. The classification is based on the concept of the collective Hamiltonian in the $\gamma$ degree of freedom, where
 "soft" and "rigid" refer to the potential confining the fluctuations of the shape around the mean value. We discussed this
 classification scheme in detail in Ref. \cite{Rouoof2024}. Accordingly,  $^{72}$Ge and $^{186}$Os are "$\gamma$-soft"
 and $^{76}$Ge, $^{192}$Os and $^{194}$Pt are "$\gamma$-rigid", while as
 $^{188}$Os and $^{190}$Os indicate the transitional behaviour between soft- and rigid-motion.  $^{168}$Er displays
 very small $S(I)$ values, which indicate a small deviation from the axial shape. 
% Fig. \ref{S-gg} shows the staggering parameter of the $\gamma$ $\gamma$ band.

%===========fig4========================================
%\begin{figure}[htb]
% \centerline{\includegraphics[trim=0cm 0cm 0cm
%0cm,width=0.5\textwidth,clip]{stag2.pdf}} \caption{(Color
%online) Staggering parameters $S(I)$ of the $\gamma\gamma$-band in $^{72,76}$Ge, $^{164,166,168,170}$Er,$^{186,188,190,192}$Os and
%$^{194}$Pt.  {\bf add the experimental S(I). Are the 194PT values the most recent ones?}
%  }
%\label{S-gg}
%\end{figure}
%==================================================

%===========fig5========================================
\begin{figure}[htb]
 \centerline{\includegraphics[trim=0cm 0cm 0cm
0cm,width=0.5\textwidth,clip]{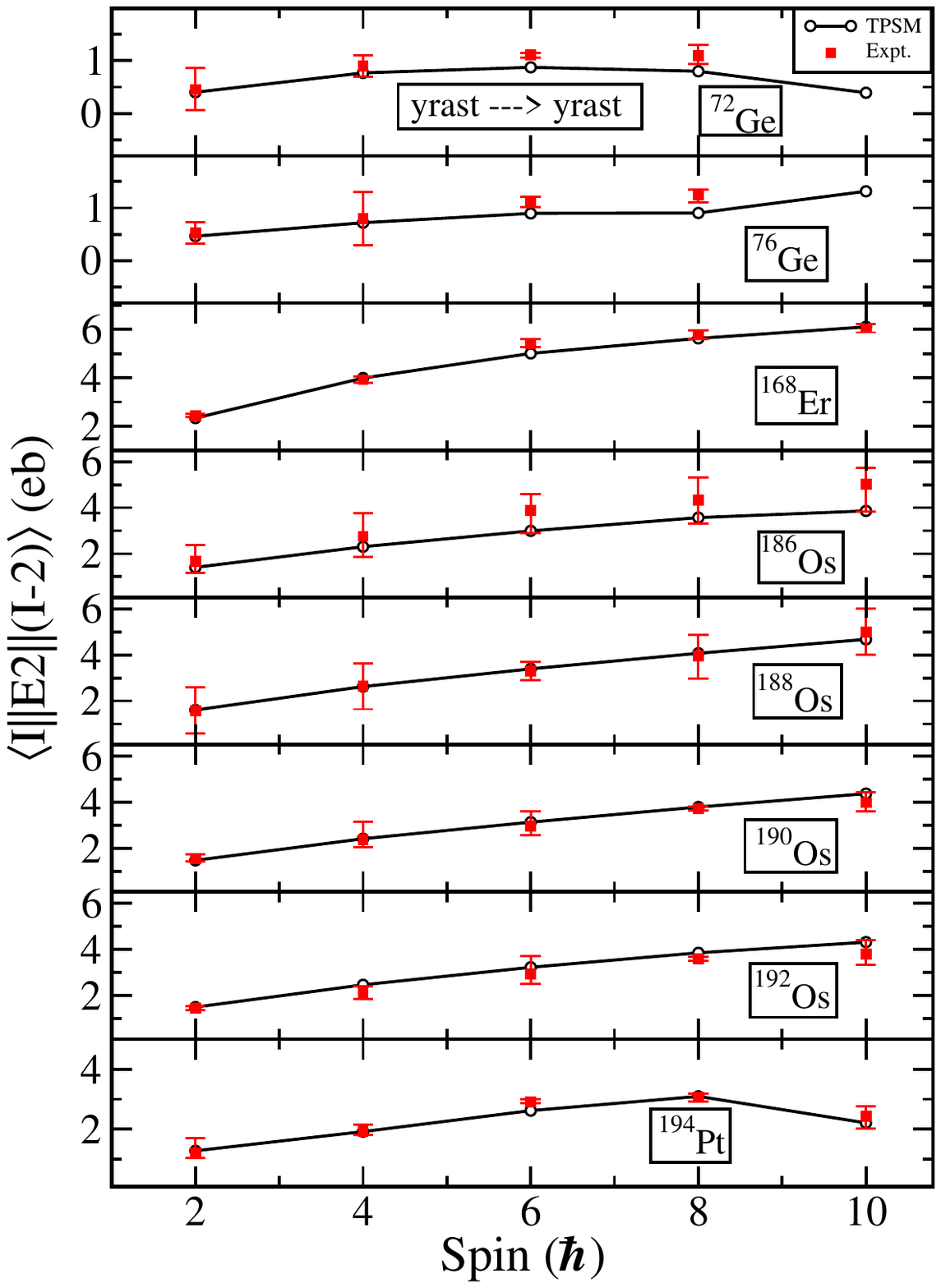}} \caption{(Color
online) Reduced  in-band E2 matrix elements for transitions yrast $\rightarrow$ yrast in $^{72,76}$Ge, $^{168}$Er,$^{186,188,190,192}$Os and
$^{194}$Pt. Expt. data is taken from the Refs. \cite{Ayangeakaa2016,Ayangeakaa2019,Ayangeakaa2023,Kotlinski1990,Wu1996,Allmond2008}.
  }
\label{E22-in-y}
\end{figure}
%===================================================
%===========fig6========================================
\begin{figure}[htb]
 \centerline{\includegraphics[trim=0cm 0cm 0cm
0cm,width=0.5\textwidth,clip]{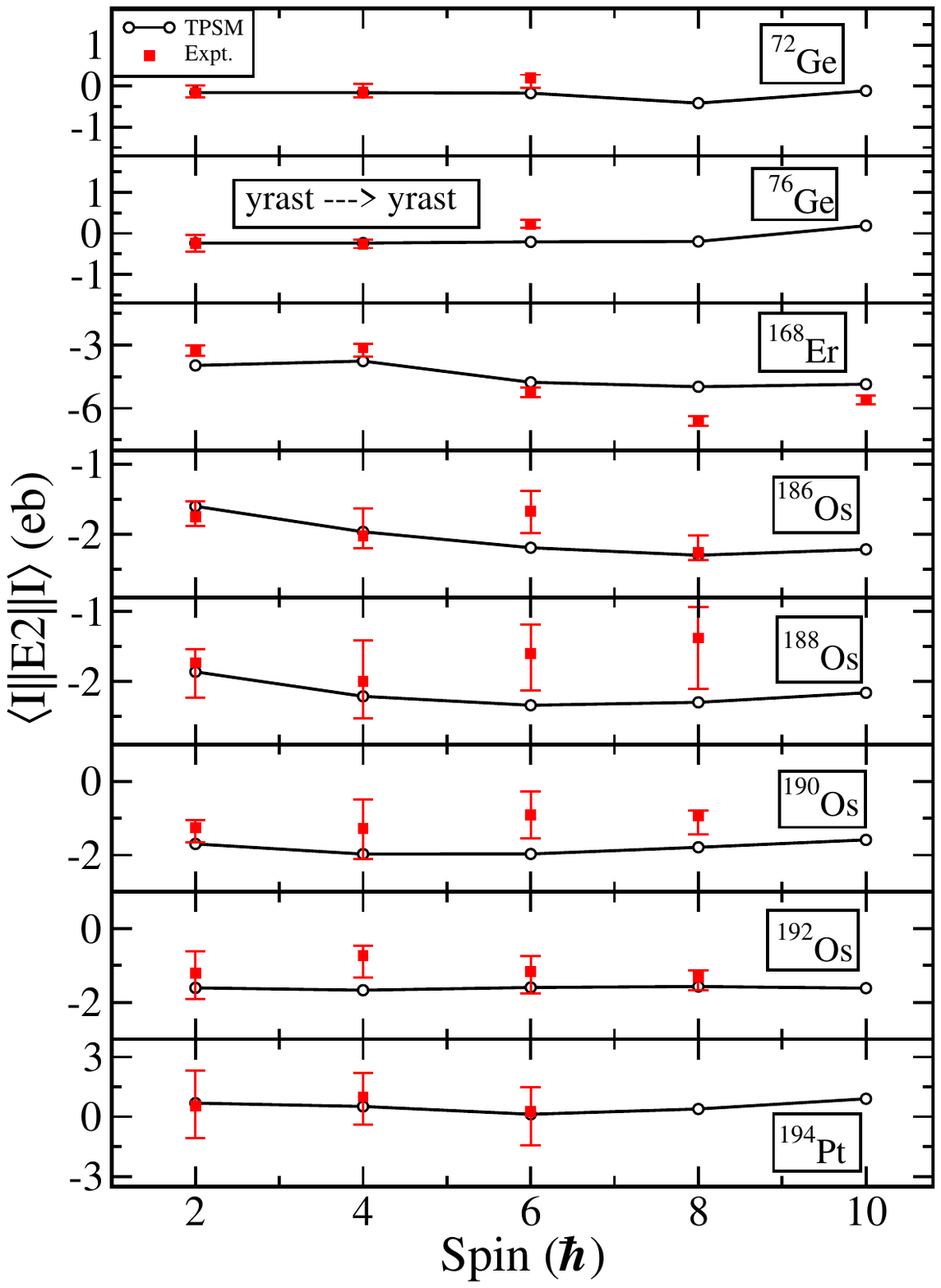}} \caption{(Color
online) Reduced diagonal E2 matrix elements for the yrast states in $^{72,76}$Ge, $^{168}$Er,$^{186,188,190,192}$Os and
$^{194}$Pt. Expt. data is taken from the Refs. \cite{Ayangeakaa2016,Ayangeakaa2019,Ayangeakaa2023,Kotlinski1990,Wu1996,Allmond2008}.
  }
\label{E2-dia-y}
\end{figure}
%===================================================  

%===========fig7========================================
\begin{figure}[t]
 \centerline{\includegraphics[trim=0cm 0cm 0cm
0cm,width=0.5\textwidth,clip]{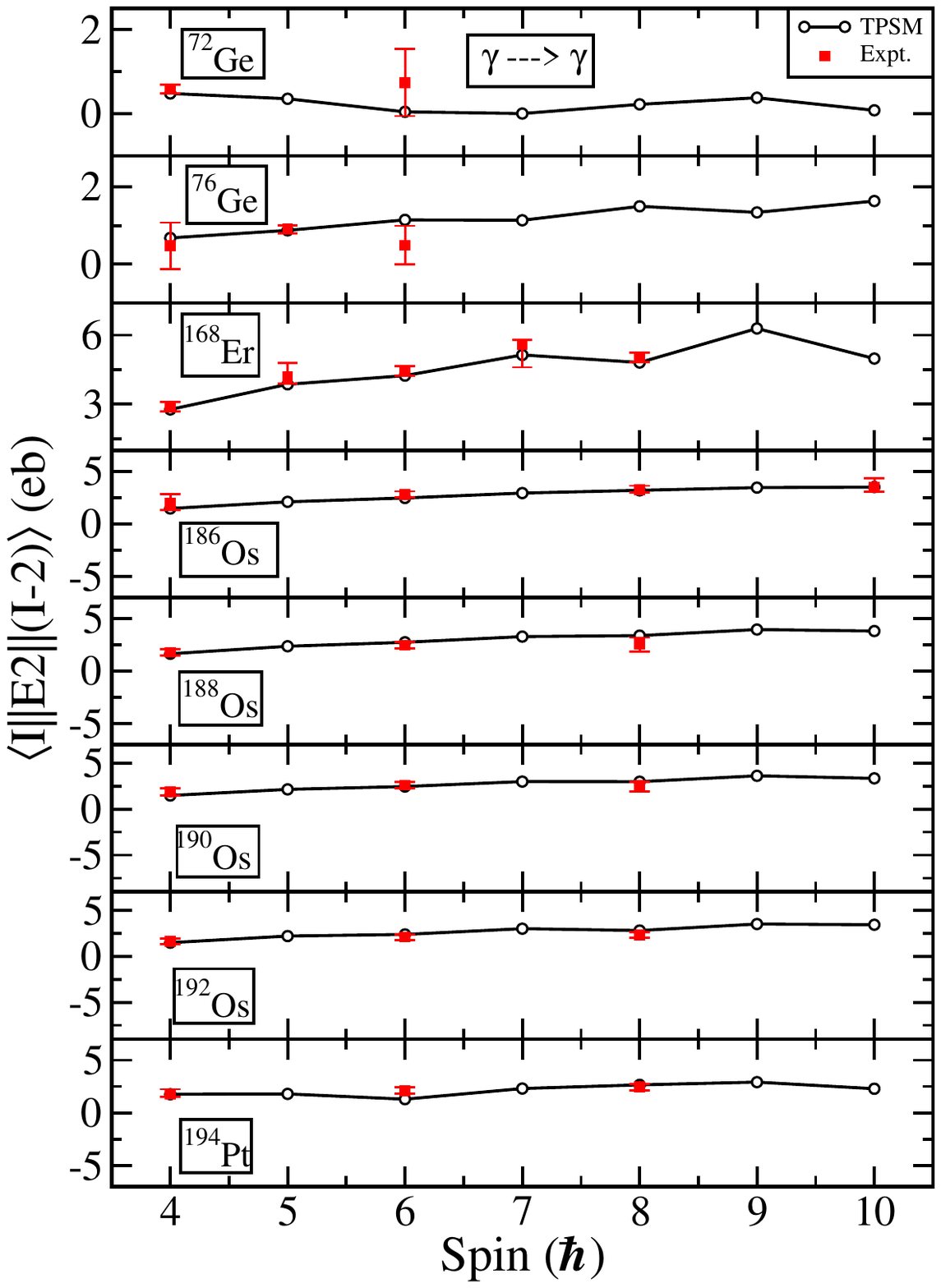}} \caption{(Color
online) Reduced in-band E2 matrix elements for transitions $\gamma$ $\rightarrow$ $\gamma$ in $^{72,76}$Ge, $^{168}$Er,$^{186,188,190,192}$Os and
$^{194}$Pt. Expt. data is taken from the Refs. \cite{Ayangeakaa2016,Ayangeakaa2019,Ayangeakaa2023,Kotlinski1990,Wu1996,Allmond2008}.
  }
\label{E22-in-g}
\end{figure}
%===================================================
%===========fig8========================================
\begin{figure}[t]
 \centerline{\includegraphics[trim=0cm 0cm 0cm
0cm,width=0.5\textwidth,clip]{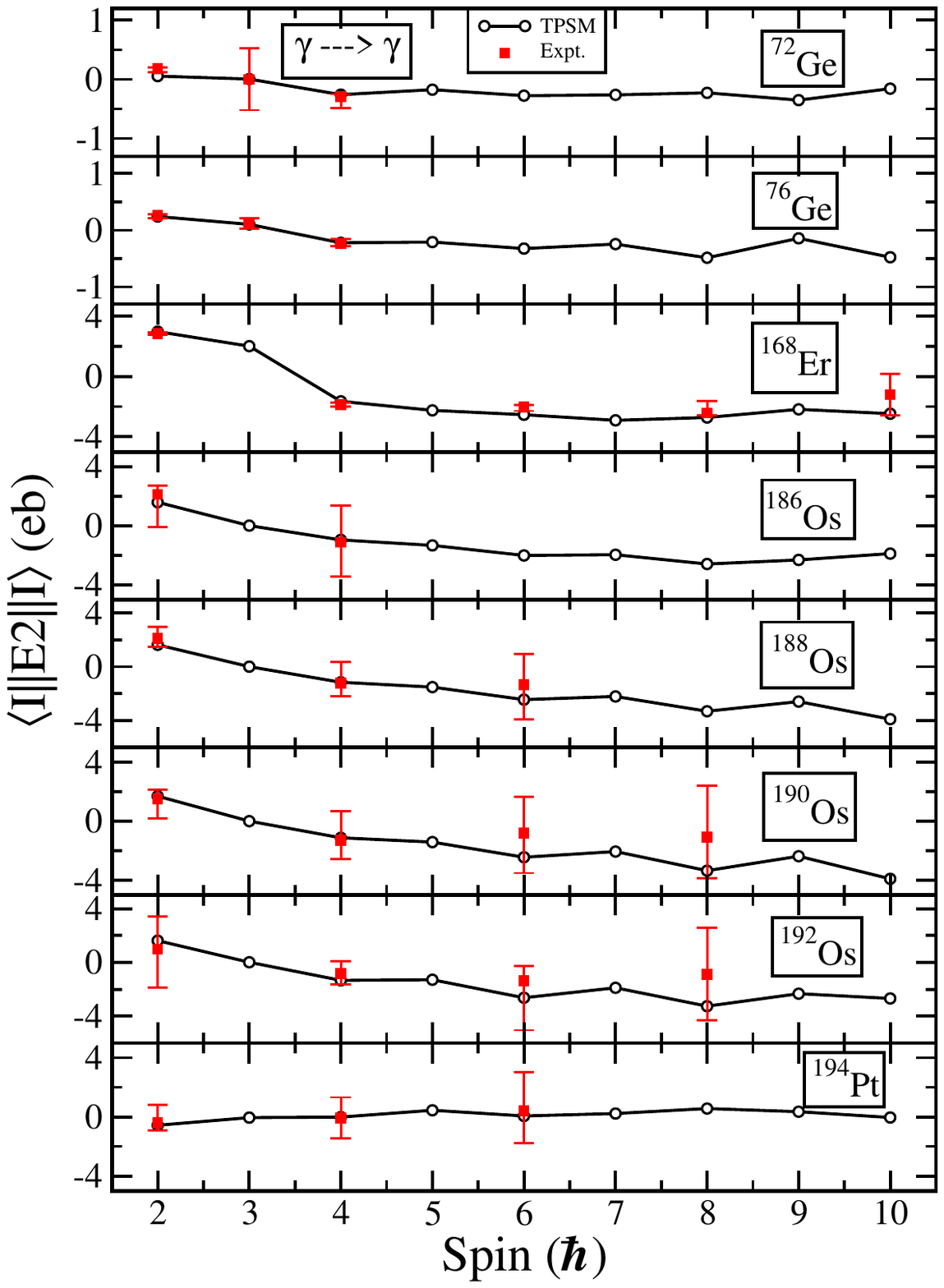}} \caption{(Color
online) Reduced diagonal E2 matrix elements  for the $\gamma$-band states $^{72,76}$Ge, $^{168}$Er,$^{186,188,190,192}$Os and
$^{194}$Pt. Expt. data is taken from the Refs. \cite{Ayangeakaa2016,Ayangeakaa2019,Ayangeakaa2023,Kotlinski1990,Wu1996,Allmond2008}.
  }
\label{E2-dia-g}
\end{figure}
%===================================================
%===========fig9========================================
\begin{figure}[htb]
 \centerline{\includegraphics[trim=0cm 0cm 0cm
0cm,width=0.5\textwidth,clip]{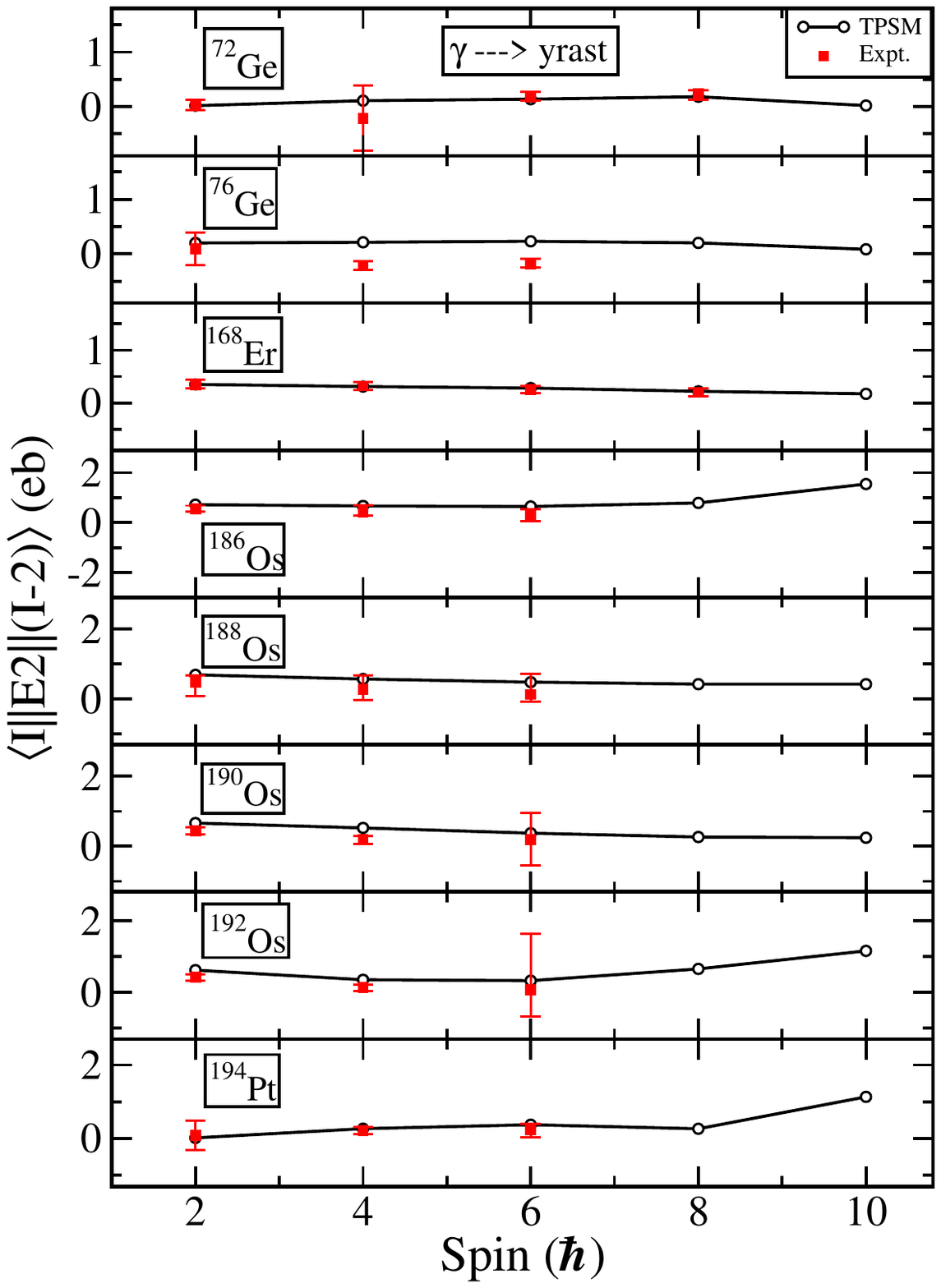}} \caption{(Color
online) Reduced inter-band E2 matrix elements for transitions $\gamma$ $\rightarrow$ yrast in $^{72,76}$Ge, $^{168}$Er,$^{186,188,190,192}$Os and
$^{194}$Pt. Expt. data is taken from the Refs. \cite{Ayangeakaa2016,Ayangeakaa2019,Ayangeakaa2023,Kotlinski1990,Wu1996,Allmond2008}.
  }
\label{E22-gy}
\end{figure}
%===================================================
%===========fig10========================================
\begin{figure}[htb]
 \centerline{\includegraphics[trim=0cm 0cm 0cm
0cm,width=0.5\textwidth,clip]{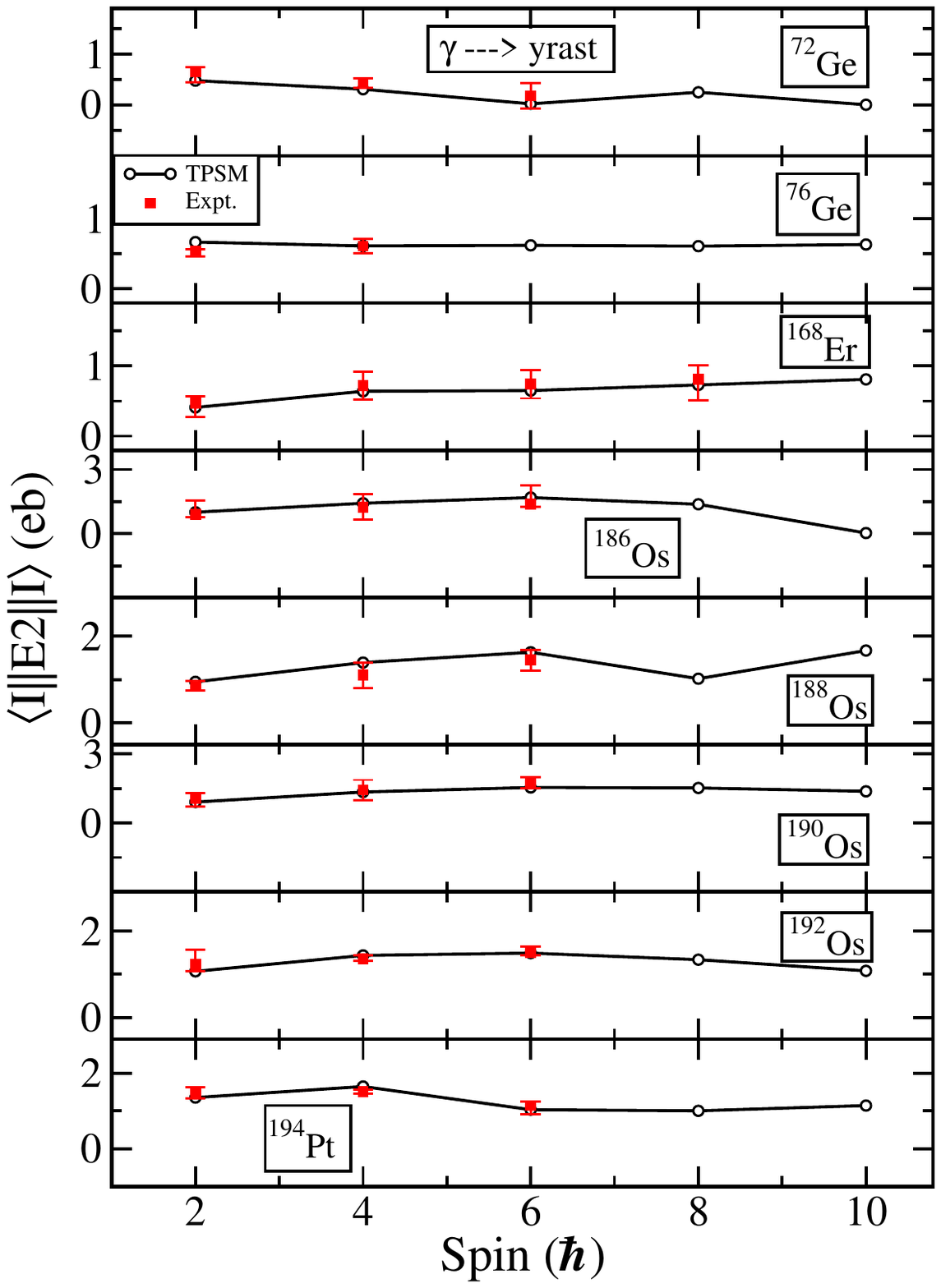}} \caption{(Color
online) Reduced inter-band E2 matrix elements for transitions $\gamma$ $\rightarrow$ yrast in $^{72,76}$Ge, $^{168}$Er,$^{186,188,190,192}$Os and
$^{194}$Pt. Expt. data is taken from the Refs. \cite{Ayangeakaa2016,Ayangeakaa2019,Ayangeakaa2023,Kotlinski1990,Wu1996,Allmond2008}.
  }
\label{E20-gy}
\end{figure}
%===================================================

\subsection{$E2$ matrix elements}

In this section, we provide a detailed comparison of the $E2$ matrix elements 
for  $^{72}$Ge, $^{76}$Ge, $^{168}$Er, $^{186}$Os, $^{188}$Os, $^{190}$Os,
$^{192}$Os and  $^{194}$Pt evaluated using the TPSM approach  with
the  available experimental values. These eight nuclides
have been chosen because large sets of $E2$ matrix elements have been measured 
through COULEX
experiments. Partial sets of the $E2$  matrix elements for yrast- and $\gamma$-bands are displayed in Figs. \ref{E22-in-y}-\ref{E20-gy}
and all the matrix elements up to $I=10$ for yrast-, $\gamma$- $\gamma\gamma$- and $0_2^+$-bands are listed in
Tables \ref{E2INBD} to \ref{E2INTERBS3} of the Appendix \ref{E2AP}. The total number 
of 1496 tabulated matrix elements listed in the tables should
be useful for future experimental and theoretical comparisons. We have limited the evaluation of the
matrix elements to $I=10$ as most of the
COULEX data are  restricted this spin value.

%As the sign  of an eigenstate is indeterminate,  the signs of some transition  transition matrix elements  must be determined by a phase convention. The one
%adopted in present manuscript work is provided in Table \ref{sign}, which  is similar to that used
%in most of the experimental studies of $E2$ matrix elements \cite{Kotlinski1990,Ayangeakaa2016,Wu1996,104Ru}. 
%The signs of the remaining matrix elements are fixed by the phase convention. 
%The resulting TPSM $E2$ matrix elements are listed in Tables \ref{E2INBD} - \ref{E2INTERBS3} and compared with available experimental values.
 
Before proceeding further with the discussion of the matrix elements, we would like to note that there is a partial
ambiguity in the signs of the non-diagonal reduced  matrix elements because the sign of the eigenvector does not have a physical
relevance. It is a matter of the adopted phase convention. Following the standard convention in evaluating the $E2$ matrix elements
in COULEX experiments  \cite{Kotlinski1990,Ayangeakaa2016,Wu1996,104Ru}, 
 we assign a $+$ sign to all in-band matrix elements $\langle I\vert\vert E2\vert\vert I-2\rangle$. Further, one needs to
 assign the sign of one 
 matrix element $\langle I\vert\vert E2\vert\vert I-1\rangle$ between the even and odd-$I$ sequences of the
 $\gamma$- and $\gamma\gamma$-bands 
 and the sign of one matrix element  between the bands.  This fixes the signs of the eigenvectors and the signs of the
 remaining matrix elements. 
 [The situation is analogous to the transition energies. Knowing the in-band transition
 energies and one transition energy between each band fixes all the
 transition energies between the bands.] Figs. \ref{E22-in-y}-\ref{E20-gy} depict the matrix elements
 with the sign convention of the COULEX experiments.
In the event this sign convention required reversing the sign of the matrix element
obtained in the TPSM calculations,  the original sign is quoted in the parentheses of  the matrix element listed in the table.

The reduced $E2$ matrix elements corresponding to the transitions $I \rightarrow I-2$ within the yrast-band 
are depicted in Fig. \ref{E22-in-y} and listed in Table \ref{E2INBS1}.
 According to the adopted phase convention they are positive.
 The matrix elements exhibit an increasing trend with spin, which
is a general feature obtained in most of models \cite{Kotlinski1990,Ayangeakaa2016,Wu1996}. 
It is evident from the figure  that agreement between the
TPSM values and the measured matrix elements is quite good. 

The diagonal matrix elements of the yrast-bands, which represent  the static quadrupole moments,  are shown in Fig. \ref{E2-dia-y}
and listed in Table \ref{E2INBD}.
The signs of these matrix elements do not depend on the phase convention and represent a physical quantity.
The TPSM values reproduce the measured matrix elements quite well.  
As discussed in detail in Ref. \cite{Rouoof2024}, the negative values for the yrast states in $^{168}$Er and $^{186-192}$Os indicate a preference for prolate deformation. This is also evident from the change of the sign of the matrix elements
  from positive to negative for the values of the $\gamma$-band. 
The small
absolute values for $^{72,76}$Ge and $^{194}$Pt are evidence for the  
triaxiality being close to its maximum, where the negative sign for
the Germanium isotopes indicate 
that the shape is slightly on the prolate side where 
the moments of inertia, ${\cal J}_l<{\cal J}_s$. The positive sign for Pt indicates
that the shape is slightly on the oblate side where ${\cal J}_s<{\cal J}_l$. 
A detailed discussion
on the relationship between the moment of inertia and the $E2$ matrix elements is given in Ref. \cite{Rouoof2024}.

The stretched in-band $E2$ matrix elements for the $\gamma$-band are depicted in
Fig. \ref{E22-in-g} and displayed in Table \ref{E2INBS1}. They are positive
due  to the adopted phase convention. 
For $^{72}$Ge, the values are almost constant with spin,
and for other nuclides a slightly increasing trend is noted, which  is consistent with the experimental data. 
The small values for $^{72,76}$Ge and $^{194}$Pt are evidences for  their strong triaxiality \cite{Rouoof2024}. 

The TPSM diagonal matrix elements of the $\gamma$-band are
shown in Fig. \ref{E2-dia-g} and listed in Table \ref{E2INBD}. For the $I=2^+$ and $3^+$ states of  $^{168}$Er and $^{186-192}$Os
they are positive and change to
negative values with increasing $I$, which indicates  a preference for prolate deformation. The measured
values for the known states
also display this feature, although in some cases the error bars are too large for a definite comparison. 
The small absolute values for $^{72,76}$Ge and $^{194}$Pt are further evidence for strong triaxiality \cite{Rouoof2024}. 
Further measurements of the static quadrupole moments of the members of the $\gamma$-bands would be of considerable interest.

The matrix elements $I \rightarrow (I-2)$ for the transitions between the $\gamma$- and yrast-bands are plotted in Fig. \ref{E22-gy} and
presented in Table \ref{E2INTERBS1}.
 For  the $ I \rightarrow I$ transitions they  are displayed in
Fig. \ref{E20-gy} and  listed in  Table \ref{E2INTERBS2}. These matrix elements display a smooth dependence on
 spin. The TPSM  values are in good agreement with the measured ones. It is to be noted that the signs
of the matrix elements  $\langle 2^+_2|| E2) || 0^+_1\rangle $ and $\langle 2^+_2||  E2) || 2^+_1\rangle $ 
are fixed by the phase convention. The reduced matrix elements for the transitions from the odd-$I$ members of the $\gamma$-band 
to the yrast-band  are listed in  Table \ref{E2INTERBS2}. 
The TPSM  values deviate from the measured two values for  $^{76}$Ge.

Table \ref{E2INBD} displays the diagonal TPSM matrix elements  for
the $\gamma\gamma$-bands. The measured values for the  four Osmium  isotopes deviate notably from the TPSM results.
Table \ref{E2INTERBS1} lists the reduced matrix elements for the $I \rightarrow I-2$ transition from the 
$\gamma\gamma$-bands to the $\gamma$- and yrast-bands. 
The TPSM values again deviate from the experimental values for the Osmium isotopes.
However, the TPSM calculations show the expected enhancement of the former as compared to the latter.
An exception  is the $4_3 \rightarrow 2_2$ matrix element in $^{186}$Os, probably the state is a two-quasiparticle excitation, 
and the $\gamma\gamma$-excitation is the $4_4$ state. 

For almost all  of the $I \rightarrow I-2$ transitions connecting  the $0^+_2$-bands with  the  $\gamma\gamma$-, $\gamma$-
and the yrast-bands,
there are no experimental values to compare with the TPSM values. The measured matrix elements $\langle2^+_3||E2||0^+_1\rangle$ 
for $^{72,76}$Ge are a factor 2-3 larger than the TPSM values. 

There could be several reasons for the descrepancies between the measured values and TPSM predictions noted above and we
highlight here a few of them. 
In general, one expects that the admixture of  multi quasiparticle configurations included in the present work
should account for mean-field changes up
to a certain extent. However, substantial changes involve the admixture of a large set of
highly excited multi-quasiparticle configurations,
which are outside the  adopted basis set.  Further, the  excited $0_2^+$-band may have a shape coexistence character rather
than the two-quasiparticle type considered in the present work. The pairing gaps of the two-quasiparticle
states have been reduced  compared to the ground state band
so as to reproduce the band head energy correctly. In an accurate treatment, the quasiparticle basis states belonging
to different shapes need to be admixed within  the generator coordinate method.
The deviations of the TPSM values for the 
$\gamma\gamma$-bands in the Osmium isotopes may be caused by an incorrect description of the fragmentation
of the $\gamma\gamma$ strength, which is sensitive to the details of the fragment energies. 

%===========fig11========================================
\begin{figure}[htp]
  \centerline{\includegraphics[trim=0cm 0cm 0cm
0cm,width=0.5\textwidth,clip]{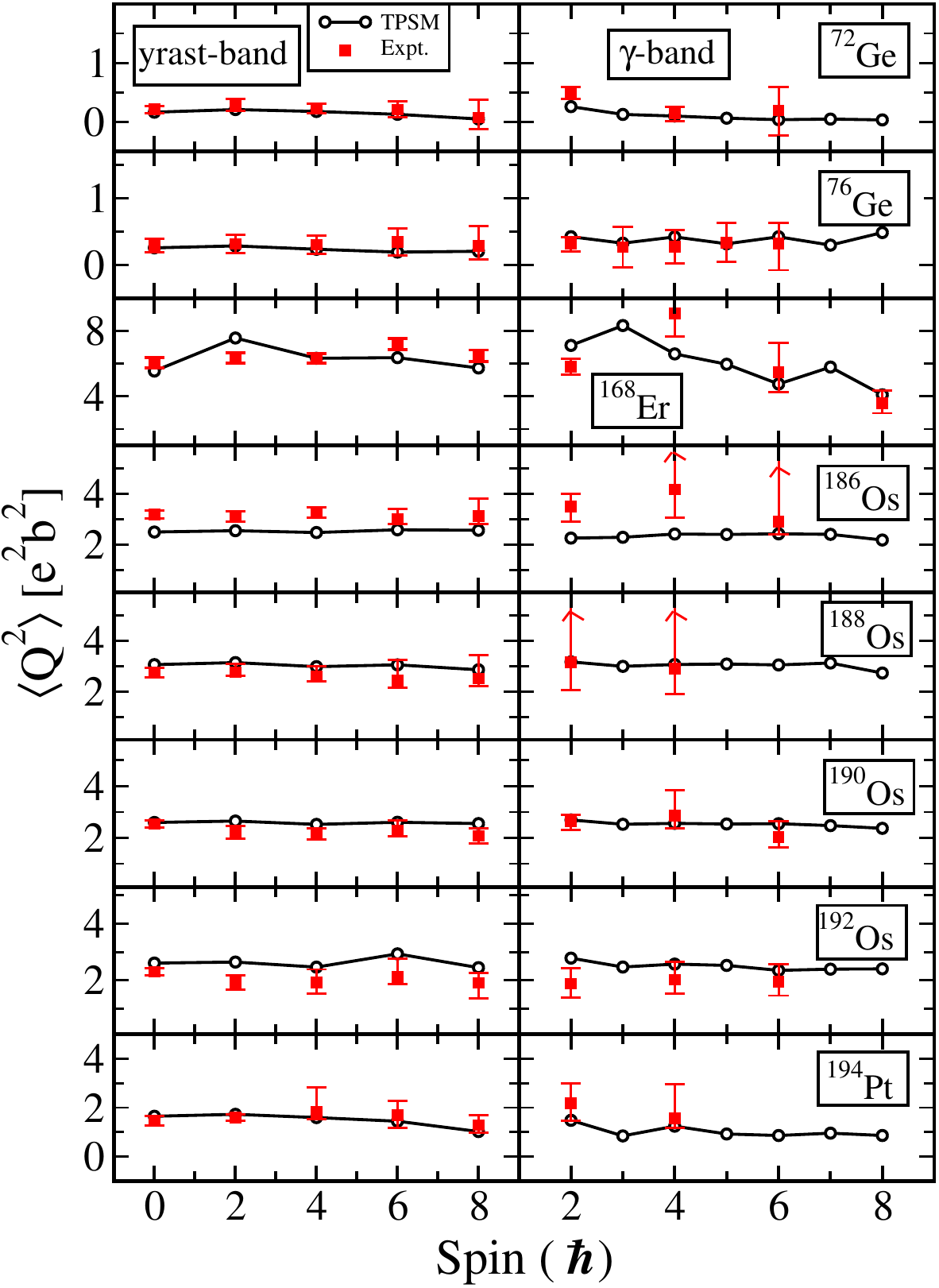}} \caption{(Color
online) Centroid $\langle Q^2 \rangle$ of the yrast- and $\gamma$-bands for $^{72,76}$Ge, $^{168}$Er,$^{186,188,190,192}$Os and
$^{194}$Pt isotopes. %The \mbox{TPSM 1} are calculated with the set of 187 $E2$ matrix elements. The \mbox{TPSM 2} points are calculated with the same restricted set of $E2$ matrix elements that is used to calculate the experimental points from the COULEX data.
Expt. data is taken from the Refs. \cite{Ayangeakaa2016,Ayangeakaa2019,Kotlinski1990,Wu1996}.
  }
\label{f:Q2}
\end{figure}
%===================================================
%===========fig12========================================
\begin{figure}[htp]
  \centerline{\includegraphics[trim=0cm 0cm 0cm
0cm,width=0.5\textwidth,clip]{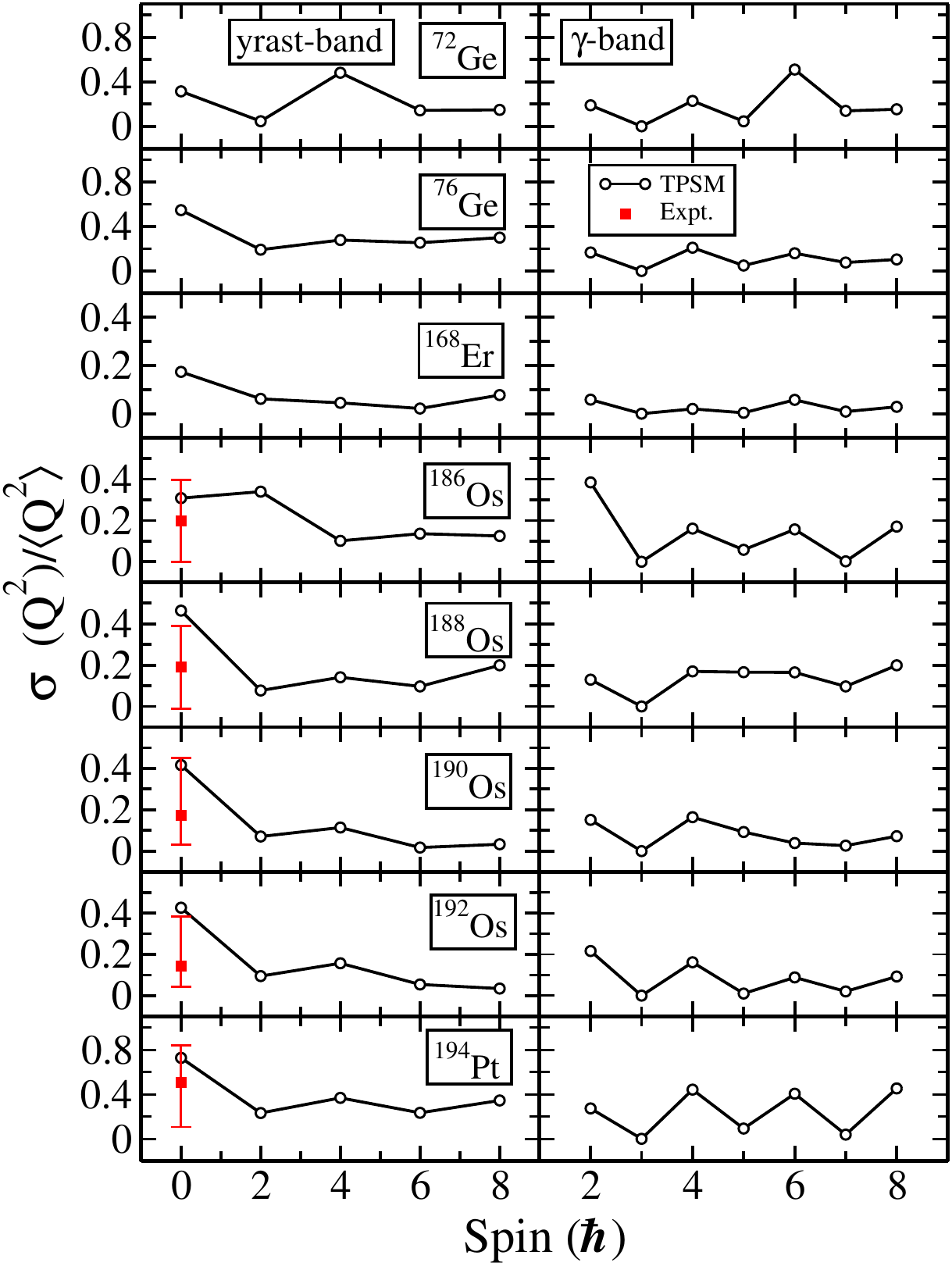}} \caption{(Color
online) Dispersion in $ Q^2$ of the yrast- and $\gamma$-bands for $^{72,76}$Ge, $^{168}$Er,$^{186,188,190,192}$Os and
$^{194}$Pt isotopes.%The \mbox{TPSM 1} are calculated with the set of 187 $E2$ matrix elements. The \mbox{TPSM 2} points are calculated with the same restricted set of $E2$ matrix elements that is used to calculate the experimental points from the COULEX data. 
 Expt. data is taken from the Refs. \cite{Ayangeakaa2016,Ayangeakaa2019,Kotlinski1990,Wu1996}. }
\label{f:sigmaQ2}
\end{figure}
%=================================================== 

Table \ref{E2INBS1} provides the in-band stretched transition $E2(I \rightarrow \\ (I-2))$ matrix elements 
for the $\gamma\gamma$-bands. There is no data available
for any of the nuclides. For the excited $0_2^+$-bands  only one transition in $^{72}$Ge is available.
The measured experimental value indicates a  deformation that is reduced by a factor of 0.73 
as compared to the deformation used in the TPSM calculations.

The $I \rightarrow (I-1)$ in-band transitions are listed in Table \ref{E2INBS2} for 
the $\gamma$-  and $\gamma\gamma$-bands.
Evidently, TPSM values  are in good agreement with the measured ones. 
There are no measured  intra-band matrix elements 
for the $\gamma\gamma$-bands to compare with the TPSM  values.

Table \ref{E2INTERBS1} provides the $I \rightarrow (I-2)$  inter-band $E2$ matrix elements for the
 transitions  $\gamma\gamma\rightarrow$
yrast, $\gamma\gamma\rightarrow\gamma$,  $0_2^+\rightarrow$  yrast, $0_2^+\rightarrow\gamma$ and 
excited $0^+ \rightarrow\gamma\gamma$.
 The    $\gamma\rightarrow$  yrast matrix elements  are measured  for all the
eight nuclei studied, at least, up to $I=6$.
The TPSM values agree within the error bars
with the corresponding measured values.

The $I \rightarrow (I-1)$ inter-band transitions matrix elements  are listed in Table \ref{E2INTERBS2}. Only the matrix elements  between the $\gamma$-
and yrast-bands in $^{76}$Ge and the matrix  elements between the $\gamma$- and $\gamma\gamma$-bands in the Osmium isotopes are measured,
which deviate from the TPSM values. The discrepancy for $^{186}$Os has already been attributed to the two-quasiparticle nature of the $4^+_3$ TPSM state in contrast to its
 $\gamma\gamma$ nature in experimental data.

Table \ref{E2INTERBS30} provides the $I \rightarrow (I+1)$ inter band matrix elements. The TPSM reasonably accounts for 
the measured $\gamma\rightarrow$ yrast values. There are no experimental matrix elements available for the other transitions.   
More data are required to
 better analyze the TPSM predictions.

Table \ref{E2INTERBS3} quotes the $I \rightarrow (I+2)$ inter-band matrix elements and
between the yrast- and $\gamma$-band, which are reasonably well reproduced. 
The TPSM matrix elements from the  $\gamma\gamma$ to the yrast-band and to the $\gamma$-band show significant deviations 
from the few observed values. As already mentioned, the reason for these deviations might be attributed to the incorrect description of the
fragmentation of the  collective $\gamma\gamma$-excitation.
The matrix element for transitions between the $0^+_2$-band, and the $\gamma$- and yrast-bands 
 are provided in the tables as well.
 % and we have calculated these transitions as for some cases measured values are
%known. Again descrepancies are noted for some cases, in particular, the signs for some transitions are wrongly predicted by the
%TPSM approach. However, it is to be noted that these transitions are quite small, and a modest admixing from the
%neglected configurations could easily alter the phase of the matrix element.

%===========fig13========================================
\begin{figure}[htb]
 \centerline{\includegraphics[trim=0cm 0cm 0cm
0cm,width=0.5\textwidth,clip]{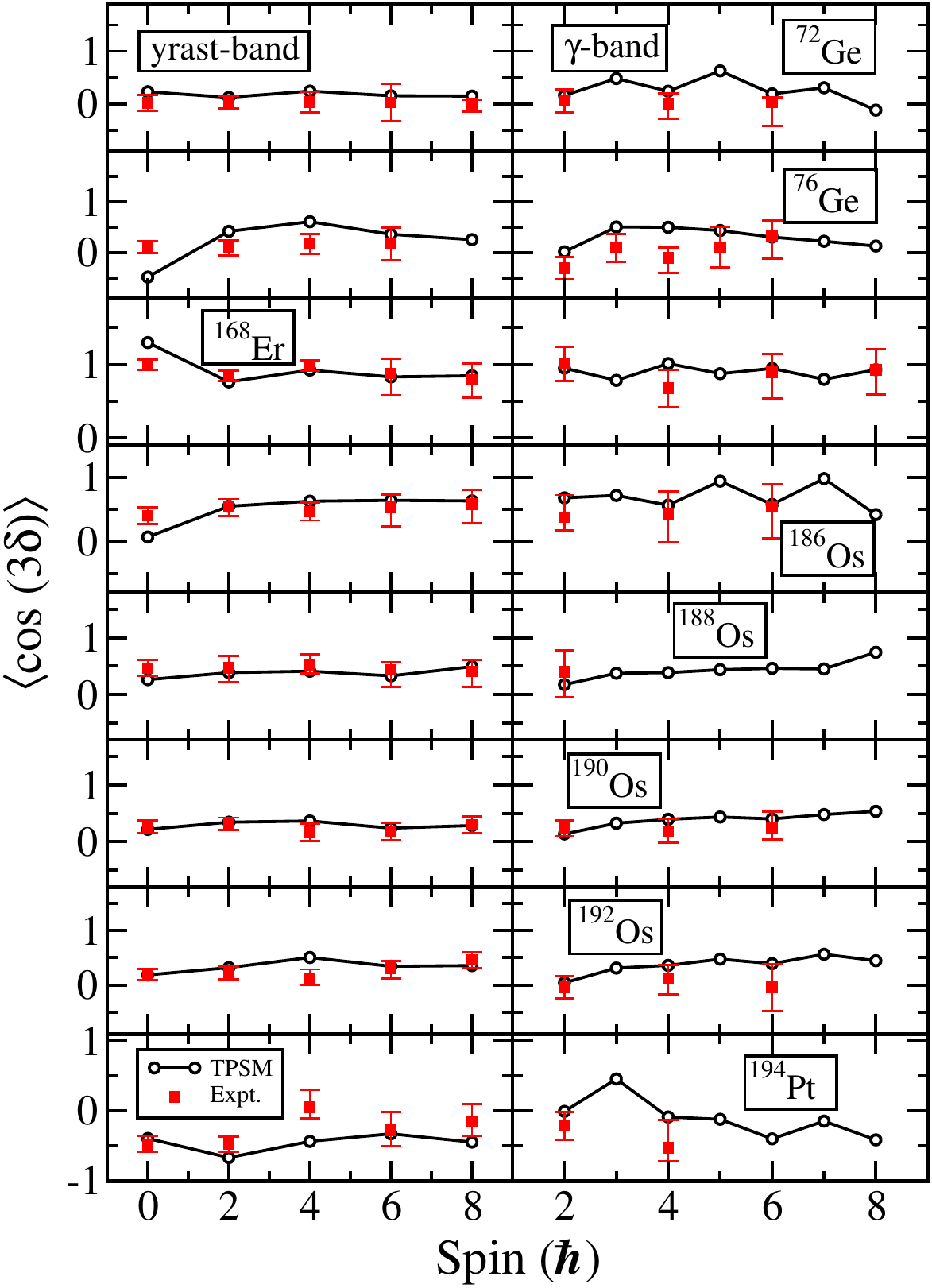}} \caption{(Color
online) Centroid $\langle \textrm{cos}3\delta \rangle$ of the yrast- and $\gamma$-bands for $^{72,76}$Ge, $^{168}$Er, $^{186,188,190,192}$Os and
$^{194}$Pt isotopes. %The \mbox{TPSM 1} are calculated with the set of 187 $E2$ matrix elements. The \mbox{TPSM 2} points are calculated with the same restricted set of $E2$ matrix elements that is used to calculate the experimental points from the COULEX data. 
 Expt. data is taken from the Refs. \cite{Ayangeakaa2016,Ayangeakaa2019,Kotlinski1990,Wu1996}. }
\label{f:cosdelta}
\end{figure}
%===================================================
%===========fig14========================================
\begin{figure}[htb]
 \centerline{\includegraphics[trim=0cm 0cm 0cm
0cm,width=0.5\textwidth,clip]{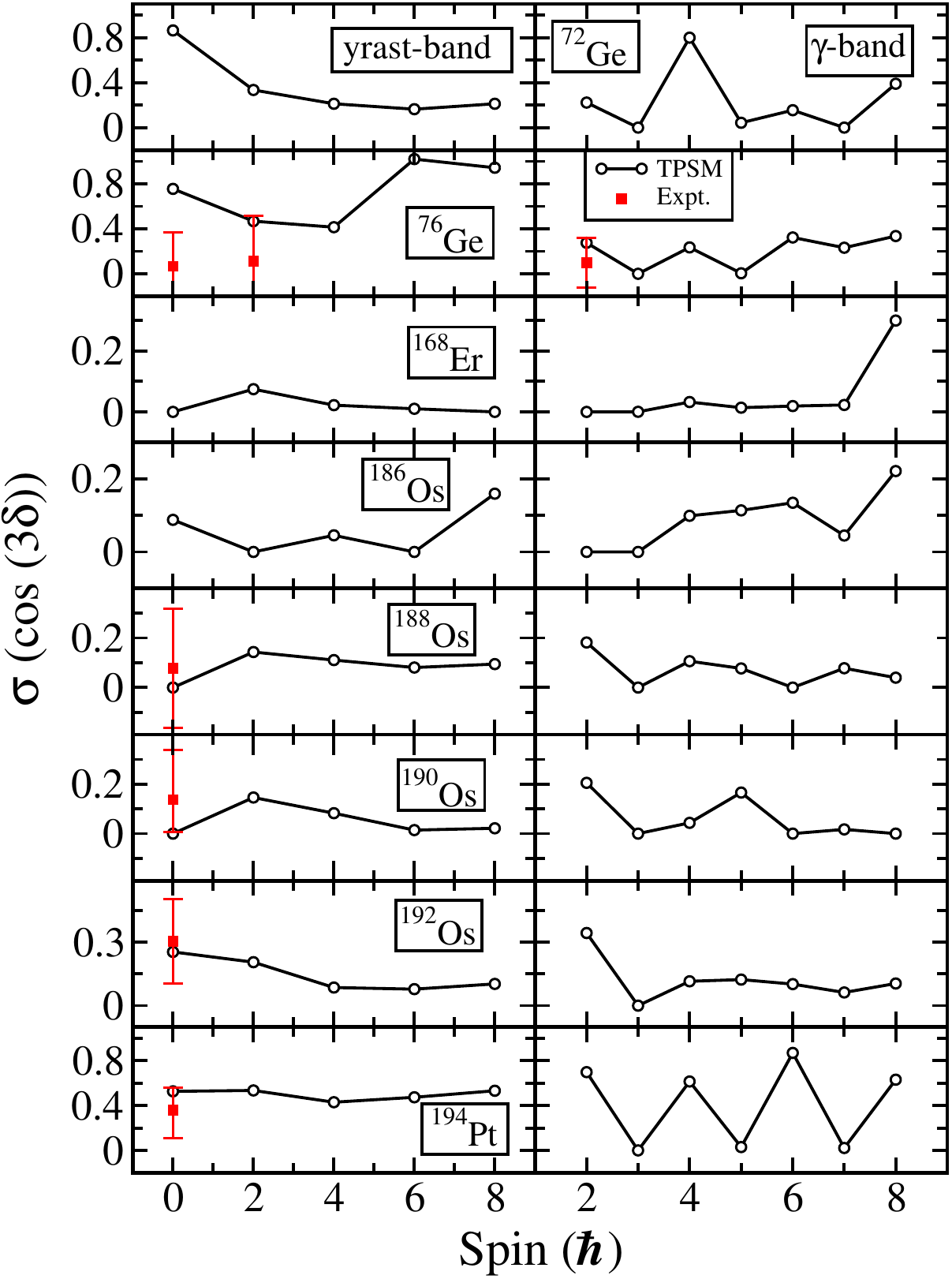}} \caption{(Color
online) Dispersion in $\textrm {cos}3\delta$ of the yrast- and $\gamma$-bands for $^{72,76}$Ge, $^{168}$Er,$^{186,188,190,192}$Os and
$^{194}$Pt isotopes. %The \mbox{TPSM 1} are calculated with the set of 187 $E2$ matrix elements. The \mbox{TPSM 2} points are calculated with the same restricted set of $E2$ matrix elements that is used to calculate the experimental points from the COULEX data. 
 Expt. data is taken from the Refs. \cite{Ayangeakaa2016,Ayangeakaa2019,Kotlinski1990,Wu1996}. }
\label{f:sigmacosdelta}
\end{figure}

\subsection{Quadrupole shape invariants} 

The quadrupole shape invariants have been evaluated using the coupled channel
least-squares search code, GOSIA \cite{GOSIA2012}.
These quantities provide model-independent information about the shape of the system and are evaluated
using the $E2$ matrix elements deduced from Coulomb excitation experiments \cite{D.Cline.Annu.Rev.Nucl.Sci,Wu1996,Kotlinski1990,Ayangeakaa2016,Ayangeakaa2019,Ayangeakaa2023}.
The shape-invariants analysis is possible
for all the eight nuclei studied in the present work as large sets of $E2$ matrix elements have been obtained  using
COULEX experiments. As a matter of fact, the availability of the large set of data for these isotopes is the reason that we decided to
perform a detailed analysis for these nuclides.

%In principle, the evaluation of a shape invariant requires all the $E2$ matrix elements 
%involved in the multiple sums, which become quite large for the higher powers of the tensor products.
%In practice, on the experimental side one is forced to truncate the sums to the matrix elements that are measured in the COULEX
%experiments. The errors by truncating the sums are expected to increase with the power of the tensor product. In order to
%probe these truncation errors,
%we have also evaluated the shape invariants with the $E2$ matrix elements calculated from the TPSM wavefunctions, and shall compare with
%experimentally deduced quantities. We have carried out the GOSIA analysis with all the matrix elements of
%yrast-, $\gamma$-, $\gamma\gamma$-
%and $0^+_2$-bands up to I=10, which are 187 in number. 
%Obviously, there are many more matrix elements
%in the TPSM basis, but we have limited up to
%I=10 for the four bands since TPSM results become inaccurate for high-spin states. The problem is that in the TPSM approach, the
%mean-field is held fixed for all the configurations, and any major modifications in the mean-field deformation and pairing
%with spin or excitation energy are not considered. As already stated in the discussion of the $E2$ matrix elements that although
%multi-quasiparticle states are considered in the model space, but these will only
%account for minor modifications in the mean-field.

In principle, the evaluation of a shape invariant requires all the $E2$ matrix elements 
involved in the multiple sums, the number of which becomes quite large.
In practice, on the experimental side one is forced to truncate the sums to the matrix elements that are measured in the COULEX
experiments. We have carried out the GOSIA analysis including
all TPSM evaluated matrix elements within and between
yrast-, $\gamma$-, $\gamma\gamma$-
and $0^+_2$-bands up to $I=10$, which are 187 in number and listed in Tables \ref{E2INBD}-\ref{E2INTERBS3}. In order to investigate possible 
truncation errors, we carried out another set of calculations with only those TPSM matrix elements for which the
experimental $E2$ values have been deduced from
the COULEX data. 
In most cases the results were close enough to justify the comparison of the
experimental shape invariants with the TPSM derived quantities from the full set.
There are exceptions where the two differ and will be commented in the following.
%Figs. \ref{f:Q2} - \ref{f:sigmacosdelta} display TPSM values for the
% shape invariants, $Q^2$ and $\cos 3\delta$ and their dispersions. 
%The intermediate sums appearing in the expressions (\ref{Q2}) and (\ref{Q3}) for the  invariants $\langle Q^2\rangle$ and $\langle Q^3\cos( 3\delta)\rangle$ 
% and of $\langle Q^4\rangle$,$\langle Q^6\rangle$ and $\langle\left(Q^3\textrm{cos}3\delta\right)^2\rangle$ needed for their respective dispersions $ \sigma(Q^2)/\langle Q^2\rangle$ 
% and $\sigma(\cos( 3\delta))$ run over a complete set of states in the TPSM vector
% space. The set of 187 $E2$ matrix elements used in the present work 
% provides more accurate shape invariant values than the restricted experimental set. It contains values for odd $I$ states
% where the available experimental matrix elements are not available.
%It is noted that overall the TPSM values with the larger set have less fluctuations than the restricted set,
%which indicates that the fluctuations  are partially caused by the 
%missing experimental $E2$ matrix elements. The GOSIA results with the restricted TPSM set 
%will be mentioned in the discussion of the results of the individual systems where major deviations were noted.

In our previous publication \cite{nazira}, we performed the shape-invariants analysis for $^{104}$Ru.
It was shown that TPSM  calculated centroids and fluctuations of the quadrupole degree of freedom were in good agreement
with those derived from the measured $E2$ matrix elements. The analysis indicated
that this nucleus is $\gamma$-soft, which  was quite unexpected because TPSM approach employs a fixed $\gamma$ value
for the mean-field potential. 
One would have
expected that the resultant shape invariants to lead to a $\gamma$-rigid shape.
However, it was demonstrated that
mixing of multi-quasiparticle excitations transforms the character of the system from $\gamma$-rigid to $\gamma$-soft.
It is shown that for a few nuclei studied in the present work, the shape also transforms from
$\gamma$-rigid and $\gamma$-soft in conformity with the shape inferred from the COULEX data.

Figs. \ref{f:Q2} to \ref{f:sigmacosdelta} compare the shape invariants calculated from
the TPSM $E2$ matrix elements with those evaluated from the experimental matrix elements,
which were discussed in the previous section.
 Fig. \ref{f:Q2} displays for the yrast- and the $\gamma$-bands the quantity $\langle Q^2 \rangle$, which is a measure 
 of the deviation from the spherical shape.
 It is approximately proportional to the mean value,   $\langle\beta^2\rangle$, of the deformation parameter of the
 collective model  \cite{nazira}.
 This quantity is almost constant for the yrast- and $\gamma$-bands of all the eight nuclei studied, except for the $\gamma$-band
 in $^{168}$Er, which will be discussed below.
The dispersions $ \sigma(Q^2)/\langle Q^2\rangle$ in Fig. \ref{f:sigmaQ2} are small for the yrast-bands,
except for the Germanium isotopes. For the $\gamma$-bands,  $ \sigma(Q^2)/\langle Q^2\rangle$ shows a larger even-odd
staggering as compared to the weaker staggering in the yrast-bands. This staggering is analogous to those found for
$S(I)$ of the energies, the origin of which has been discussed in Ref. \cite{Rouoof2024}. 

Fig. \ref{f:cosdelta} displays the centroids  $\langle\cos( 3\delta)\rangle$, 
which approximately represent  the mean values of $\cos 3\gamma$ of 
the triaxiality parameter. The detailed relation is
discussed in our previous publication \cite{nazira}. It is evident from the figure that, in general, the agreement 
between the calculated and the experimental values is quite reasonable. 

Fig. \ref{f:sigmacosdelta} depicts the dispersion, $\sigma(\cos( 3\delta))$. The experimental values are available only for a few cases \cite{D.Cline.Annu.Rev.Nucl.Sci,Wu1996,Ayangeakaa2019}. The TPSM 
values are in good agreement, except for $^{76}$Ge. The TPSM calculations 
with the restricted set of $E2$ matrix elements as used 
to obtain the experimental dispersion gave a smaller value, which falls into
the range of the error bars. Hence, we attribute the discrepancy to 
the restriction of the experimental $E2$ matrix elements available from 
the COULEX experiments. It is expected that the errors caused by the 
restricted set, increase with the power of the involved tensor products,
which is highest for $\sigma(\cos( 3\delta))$.
%We have also calculated the dispersions with the
%restricted TPSM set, and the results are closer to the experimental numbers for $^{76}$Ge.
%This is expected because there are many more matrix elements involved in the sums and the
%results with a larger set gives rise to more deviations. 
In the $\gamma$-bands of several nuclei, $\sigma(\cos (3\delta))$ shows an even-odd effect
with no clear phase relation to the energy staggering $S(I)$.

We shall now discuss the TPSM results of the individual systems and compare them with earlier investigations.
The nucleus $^{72}$Ge is one among only three nuclides in the
periodic table far from closed shell that has  $0^+_2$ as the first excited state.
Its experimental energy 
staggering $S(I)$ in Fig. \ref{S-g}
displays the even-$I$-down pattern that corresponds to the $\gamma$-soft shape
in the context of the collective model.
Detailed investigation of this nucleus has been performed using COULEX data with a total of 46 $E2$ and $M1$ matrix elements
extracted \cite{Ayangeakaa2016}, 
which were then used to obtain the quadrupole shape invariants.
First of all, the TPSM calculation reproduces the even-$I$-down pattern of $S(I)$, 
though with a smaller amplitude. For the 
yrast- and $\gamma$-bands, the empirical  $\langle Q^2 \rangle$
indicates a constant deformation and $\langle \textrm{cos}(3 \delta) \rangle$ implies a constant triaxiality of 
$\gamma \approx 30^\circ$ and the TPSM results are in good agreement. 
 For the $\gamma$-band, $\langle \textrm{cos}3\delta \rangle$
displays an even-$I$-down staggering in conformity with the energy staggering $S(I)$. 
The dispersions $\sigma(\textrm{cos}(3\delta))$ indicate large fluctuations
for the ground state, which decrease with $I$ in the yrast-band.
For the $\gamma$-band, the fluctuations are small except for $I=2$.

The experimental shape invariants of Ref. \cite{Ayangeakaa2016} indicate
that the yrast- and the excited $0^+_2$-bands have
similar deformation and triaxiality parameters, and  the authors suggested that
the two bands represent an equal mixing of two different unperturbed 
triaxiality deformed shapes.
The present TPSM calculation is unable to reproduce the 
very low experimental energy of $0^+_2$ state, which might be a pairing vibration.
The calculated large dispersion, $\sigma(Q^2)/\langle Q^2\rangle$, of the $0^+_1$ state
may be an indirect evidence for the existence of such a low-lying $0^+_2$ state.

%The TPSM  calculations with the restricted set of matrix elements give a smaller value for 
%of the ground state as compared to the results with the larger set shown in Fig. \ref{f:sigmacosdelta}. However, the remaining values
%of the yrast band 
% tend to become similar with increasing $I$. This seems to indicate that
% the restricted set is unreliable to estimate
% the ground state fluctuations in the $\gamma$ degree of freedom.

The energy staggering $S(I)$ of the $\gamma$-band in $^{76}$Ge shows the even-$I$-up pattern
which in the context of the collective model 
classifies the nucleus as being one of the rare  examples of being "$\gamma$-rigid" \cite{JS21}.
However, as discussed in Ref. \cite{Rouoof2024}, the amplitude of $S(I)$ 
is far below the Davydov-Filippov limit of a rigid triaxial rotor.
For this reason the authors of Ref.  \cite{Ayangeakaa2019}
 have extensively studied this nucleus by means  of COULEX data in order to probe directly its triaxiality. 
 The measured $\gamma$-ray intensities were analyzed using the computer 
 code GOSIA, and absolute
values and signs of 103 matrix elements were determined. 

%As seen in Fig. \ref{S-g}, the TPSM very well reproduces the experimental $S(I)$ values.
For the yrast- and $\gamma$-bands, $\langle Q^2 \rangle$ in Fig. \ref{f:Q2}
indicates a constant value of deformation. 
The experimental $\langle \textrm{cos}(3\delta) \rangle$ values for the two 
 bands indicate triaxiality close to the maximum of $30^\circ$.
 The TPSM calculations suggest a gradual
transition from a slightly oblate to a slightly prolate shape for the yrast-band.
The TPSM calculations with the reduced set of matrix elements gives a more constant value, which is closer to the value derived
from the experimental $E2$ matrix elements. 
For the $\gamma$-band, the TPSM calculations give a smooth 
$I$-dependence with values that are skewed
towards the prolate side.
%Hence, the staggering of the TPSM1 seems to be
%an artifact caused by truncation of the set of $E2$ matrix elements.

Both for $^{72,76}$Ge isotopes, the TPSM calculations with the restricted set of $E2$ matrix elements give smaller values of 
$\sigma(\textrm{cos}(3\delta))$ than calculation with the full set for the $I=$ 0 and 2 yrast states. As mentioned already, the 
 TPSM values from the restricted set 
 agree well with the experimental values in $^{76}$Ge. Therefore, 
the small fluctuations of $\gamma$ around its mean values, indicated  
by the experimental values, may be an artifact, and the actual fluctuations are
substantially larger. 

%{\bf The following is an interesting observation. The S(I) is not always 
%consistent with the softness derived from E2 m.e. It should be kept.}

The  TPSM values for $I=$0, 2 and 4 of $\langle \textrm{cos}(3\delta) \rangle$
and $\sigma(\textrm{cos}(3\delta))$ 
do not differ much between $^{72}$Ge and $^{76}$Ge, indicating for 
both nuclei substantial
triaxiality and softness, which decreases with $I$. Thus, the difference between
the respective even-$I$-down and even-$I$-up pattern of the staggering of the
$\gamma$-band energies is not reflected by the $E2$ transition probabilities.
It is caused by the difference of the energy and structure of the $0^+_2$
state, the origin of which remains to be ascertained. 

The Erbium isotopes are examples of well deformed prolate nuclei.
These nuclides confirm
many predictions of the quadrupole collective
model of Bohr and Mottelson \cite{BMII}, which have been tested 
extensively in the past. 
In particular, $^{168}$Er  is one of the best 
studied nuclides using different experimental
techniques \cite{Davidson1984} and more than 36 rotational bands 
have been populated for this system. A comprehensive analysis of the band structure is given in Ref. \cite{Frau18Beyond}. 
%Figs. \ref{f:Q20Eryrga} - \ref{f:sigmacosdeltaEryrga} display the TPSM shape invariants for the neighboring Er isotopes.

As expected for a well deformed prolate nuclei, $\langle Q^2\rangle$ 
is approximately constant for the yrast-band. For the $\gamma$-band,
TPSM values are about the same as the experimental values,
except for a slight bump at $I=3$, which we could not relate to 
irregularities in the individual matrix elements of $^{168}$Er 
in Figs. \ref{f:Q2} to \ref{f:sigmacosdelta}. The dispersion 
$\sigma(Q^2)/\langle Q^2\rangle$ of the yrast-band decreases with
$I$, indicating a stabilization of the deformation. This same observation holds
for the $\gamma$-band.

The mean values $\langle\cos( 3\delta)\rangle$ are  
0.8-0.9 for the yrast- and $\gamma$-bands of $^{168}$Er, which correspond to 
root mean square values  of
$\sqrt{\langle\gamma^2\rangle}=8^\circ - 12^\circ$.
 From the perspective  of the collective model, one would expect
 larger fluctuations in the $\gamma$-band than in the yrast-band, because 
 the quantum fluctuations of the one-phonon state are larger than the  fluctuations
 of the ground state.
 However, this is not the case in the microscopic TPSM calculations, where the TPSM results are consistent with the
 experimental data for $^{168}$Er, within the 
 error bars.
 %The $\sigma(\cos (3\delta))$ {\bf yet to be discussed.}
 
According to the collective  model, the $0^+_2$ state should be 
the collective $\beta$ vibration with an enhanced $E2$ transition to 
the  $2^+_1$ state of the yrast-band. The experimental limit  for the  
absolute value of the matrix element $\langle 0^+_2\vert\vert E2\vert\vert 2^+_1\rangle$ which is below 0.2 eb  \cite{Kotlinski1990} is close to the TPSM value of 0.456 eb (see Table \ref{E2INTERBS3}) and indicates the non-collective
nature of the transitions from the excited $0^+_2$ state 
($\langle 0^+_2\vert\vert E2\vert\vert2^+_1\rangle/ \langle 0^+_1\vert\vert E2\vert\vert2^+_1\rangle=0.09$). 
The other TPSM inter-band transition matrix elements that connect $0^+_2$-band with
the $\gamma$- and yrast-bands are also of the single particle order
(see Tables \ref{E2INTERBS1}-\ref{E2INTERBS3}). 
%{\bf I discarded the next sentence. It does not make sense. Low-energy $0^+_2$ generates the even-I-low pattern of gamma soft nuclei}
%This indicates that the
%low energy of the $0^+_2$ state, which causes the "$\gamma$-rigid" pattern of
%the staggering $S(I)$ of the energy of the $\gamma$ band, has another origin than a stabilization of the triaxial shape. 

The even-even Osmium and Platinum isotopes are expected to exhibit a transition from prolate to oblate shape through triaxial intermediate shapes with increasing $N$ \cite{Delaroche10,MoellerGamma}.
In a detailed COULEX experiment, $^{186}$Os, $^{188}$Os, $^{190}$Os,
$^{192}$Os and  $^{194}$Pt have been studied, 
and large sets of $E2$ matrix elements have been deduced \cite{Wu1996}
for each of these nuclides.

The values $\langle Q^2 \rangle\approx 3(\mathrm{eb})^2$ in Fig. \ref{f:Q2}
indicate 
an $I$-independent deformation for the yrast- and $\gamma$-bands  for the Os- and Pt-nuclides, 
which reflects a moderate collectivity
of $B(E2,I\rightarrow I-2)=40-90$ Wu of the intra-band transitions. 
The TPSM calculations reproduce the experimental values reasonably well, only the $^{186}$Os values 
are slightly underestimated (see Fig. \ref{E22-in-y}). 
The dispersions $\sigma(Q^2)/\langle Q^2\rangle$ show that the fluctuations of the deformation around the mean values are
small except for the ground state. The experimental values known for the ground state have large error bars and it
is difficult to make an assessment of the predicted values.
We would like to add that calculations with the restricted TPSM set, the dispersions for the ground states are much smaller, which is 
a general observation other nuclides as well.

The values of the centroid of $\langle \textrm{cos}3\delta \rangle$ 
of the yrast-band show a gradual
 shape development with $N$ : $\delta \approx 20^\circ$ for $^{186}$Os and 
$\delta \approx -20^\circ$ for $^{194}$Pt.
The  values of the $\gamma$-band develop with $N$ in a similar manner. 
It is noted from Fig. \ref{f:cosdelta} that experimental values of $\langle \textrm{cos}3\delta \rangle$
are reproduced quite well by the TPSM calculations.
%For $^{190}$Os there  are substantial differences
%between TPSM 1 and TPSM 2. The better agreement of the experimental values derived 
%from the restricted set of $E2$ matrix elements with TPSM 2  than with TPSM 1
%is unexpected, and we do not have an explanation.
%The differences between TPSM 1 and TPSM 2 are larger for the
%dispersions $\sigma(\textrm{cos}(3\delta))$. The TPSM 2 values 
%for $^{186,188,190}$Os indicate
%that their triaxial shape with some prolate preference does not fluctuate too much,
%whereas the shape of the isotones $^{192}$Os and $^{194}$Pt fluctuates significantly.
The experimental values for dispersion are available for the $0^+_1$ ground state for $^{188,190,192}$ and
$^{194}$Pt and the TPSM values are noted to be in agreement.

Based on the phase of the TPSM energy staggering $S(I)$ 
in Fig. \ref{S-g}, the nuclides $^{186,188,190,192}$Os and $^{194}$Pt
are classified as "$\gamma$-soft, rigid, soft, rigid, rigid", respectively.
However, the TPSM  dispersions $\sigma(\textrm{cos}(3\delta)$, which  
provide direct insight into the softness of the $\gamma$ degree of freedom,
do not correlate with this pattern. The experimental data also shows this disparity. This inference demonstrates
that the correlation  between the 
phase of the staggering parameter $S(I)$ and the softness of the $\gamma$ mode,
which emerges in the eigenstates of the collective Bohr Hamiltonian 
(see Ref. \cite{Rouoof2024} and earlier work cited therein)
is not realized in the 
microscopic TPSM approach for these nuclei.

\section{Summary and conclusions}\label{SAC}

In the present work, we have performed a systematic study of the $E2$ matrix elements of 
$^{72}$Ge, $^{76}$Ge, $^{168}$Er, $^{186}$Os, $^{188}$Os, $^{190}$Os, $^{192}$Os and  $^{194}$Pt nuclides. For these
eight nuclides, extensive multi-step Coulomb excitation experiments have been performed and  large sets of
$E2$ matrix elements have been deduced, which make it possible to directly ascertain
the intrinsic shape
of the nucleus by using the Kumar-Cline shape invariant sum rule analysis \cite{D.Cline.Annu.Rev.Nucl.Sci,KM72}.
In most of the cases, the $E2$ matrix elements have been analyzed using the phenomenological collective model of
Bohr and Mottelson and also the microscopic versions of it \cite{BMII,Kotlinski1990,Ayangeakaa2016,Wu1996,Ayangeakaa2019,Ayangeakaa2023}. 
These models have been partially successful in
correlating the experimental data for yrast- and $\gamma$-bands,  and it has been possible to account for the
overall behaviour of the measured properties at low spin. 
However, these models are clearly limited in scope as
quasiparticle excitations are not included, and high-spin states cannot be investigated using these approaches. As COULEX
data is now becoming available for the intermediate- and high-spin regions, it is imperative to include quasiparticle excitations
in the model description.

Further, there has been a long-standing problem to describe 
the excited $0^+$ states observed in many nuclei
using the collective models. These states are collective in nature in 
these models and transitions to the $\gamma$-
or yrast-bands should be enhanced as compared to 
the Weisskopf estimate. However, the measured transitions are 
less than the single-particle estimate and, therefore, 
have non-collective character, in contradiction to the prediction of the Bohr-Mottelson collective model.

To elucidate the rich data on $E2$ matrix elements, 
we have employed the multi-quasiparticle TPSM approach and evaluated a large set of 187 matrix elements for
each of the eight nuclei investigated in the
present work. The advantage of the TPSM approach 
is that first of all it is
microscopic in nature with the Hamiltonian consisting
of pairing and quadrupole terms. Secondly, the
model space includes multi-quasiparticle states. In the present investigation
of even-even systems, two-neutron, two-proton and two-proton$+$two-neutron quasiparticle states have been considered
apart from the zero-quasiparticle vacuum configuration. 
The angular-momentum projected states from 
these quasiparticle configurations are then employed to diagonalize the shell model Hamiltonian. In this way, the TPSM
is a useful tool to include the correlations going beyond the mean-field. 

The TPSM approach describes the appearance of a collective $\gamma$ mode
by projecting states with different values of angular momentum projection quantum number, $K$, onto
the 3-axis of triaxial shape from the considered quasiparticle configurations.
The subsequent mixing of these basis states generates the collective excitations
due to the $\gamma$ degree of freedom.
The set of $K$=2, 4, 6, ... states projected from the triaxial vacuum configuration
generates  the collective $\gamma$-, $\gamma\gamma$-,... band  structures, 
which have the 
character of triaxial rotor states. The admixture of states with different $K$
projected from two- and four-quasiparticle configurations then account for the
deviations from rigid rotation. The TPSM configuration space includes,
$\gamma$-excitation from all the quasiparticle configurations \cite{JS16}. The
$\gamma$-bands built on two-quasiparticle bands have  been identified \cite{SJ18}. 

The in-band and inter-band
$E2$ transition matrix elements for the  yrast-, $\gamma$-, $\gamma\gamma$-, $0^+_2$-bands have been investigated in detail 
and compared with the measured values, wherever available. The $E2$ matrix elements have been listed in Tables \ref{E2INBD} to \ref{E2INTERBS3} of the appendix \ref{E2AP} up to $I=10$ because the data in most of the cases is only available below or up
this spin. These are 187 matrix elements
in number for each nucleus. However, we have evaluated
the $E2$ transition matrix elements up to $I=20$, which can be made available to the interested researchers upon request.

It is evident from the comparisons provided in the figures and tables that TPSM provides a reasonable description of the
measured transition matrix elements. A few discrepancies are noted in the inter-band
matrix elements for high-spin states.  Since these matrix elements are
retarded, small admixtures from neglected configurations could cause 
significant changes. Further, the interaction employed in the TPSM analysis is quite simplistic and a more realistic is
expected to improve the agreement with the data.

From the $E2$ matrix elements, we evaluated the Kumar-Cline quadrupole shape invariants $\langle Q^2 \rangle$,
$\langle \textrm{cos}(3\delta) \rangle$, which represent the mean value of the 
deformation and the triaxiality of the nuclear shape. In addition, we calculated
the dispersions $\sigma(Q^2)/\langle Q^2\rangle$
and $\sigma(\textrm{cos}(3\delta))$, which provide an estimate of the fluctuations
around the respective mean values, i.e., of the $\beta$ and $\gamma$-softness
of the shape. 

%Two sets of calculations have been carried out: TPSM 1 includes
%only $E2$ matrix elements for which experimental values could be determined 
%from the COULEX studies. TPSM 2 includes all 187 $E2$ matrix elements within
%and between the yrast and $\gamma$ bands. TPSM 2 represents a more accurate 
%approximation to the exact invariants, which involve sums over the $E2$
%matrix elements between all TPSM eigenstates. The error due to the truncation
%is larger for the dispersions than for the mean values and larger 
%for the $\gamma$ bands than for the yrast bands. In most cases, TPSM 1 results 
%are reasonably close to  TPSM 2 ones  such that they provide credible 
%information about the mean values of $\beta$ and $\gamma$, which holds 
%for the experimental shape invariants from COULEX as well.

From the detailed analysis of the excitation energies, $E2$ matrix elements  and quadrupole shape invariants results, it is
possible to draw the following inferences $:$

\begin{enumerate}

\item
The TPSM results reproduce very well the energies of the yrast- and $\gamma$-bands. The model accounts in detail the rotational alignment of the
high-j quasiparticles as demonstrated in the $\hbar\omega(I)$ plots. 
The even-odd staggering parameter $S(I)$ of the energies of the $\gamma$-band is accurately reproduced, in particular the change of its phase with $N$ 
in the Os-Pt isotopic chain.
Based on the collective Bohr Hamiltonian,  the
even-$I$-down pattern of $S(J)$ has been associated with a shape that is soft in the
$\gamma$ degree of freedom and the even-$I$-up pattern with a shape that is rigid
\cite{castenplb,Frauendorf15,Frau18Beyond}. The authors of Ref. \cite{Rouoof2024,nazira,Jeh2022}  explained this correlation in terms of the
repulsion between the even-$I$ states of the $\gamma$-band and the $\gamma\gamma$-band, which is low in $\gamma$-soft and high in $\gamma$-rigid nuclei.
In the case of the TPSM, the staggering is caused by the repulsion between
 even-$I$ states of the $\gamma$-band and of several $0^+$-bands built on excited 
 two-quasiparticle bands. The energy and structure of the lowest 
 two-quasiparticle bands depend on $N$, which explains the rapid change of the
 phase of $S(I)$ with $N$ found in the TPSM calculations,  in accordance with the
 experimental data.

\item
Yrast- and
$\gamma$-bands in the $^{72}$Ge (even-$I$-down) and
$^{76}$Ge (even-$I$-up) are strongly triaxial on the average 
($\langle\gamma\rangle\sim 30^\circ$) and in conformity with the experimental data.
 Based on the dispersions,
$\sigma(\textrm{cos}(3\delta))$, the $I=$0, 2, 4 yrast states  
are found to be $\gamma$-soft in both nuclei.
At higher $I$, the yrast states of $^{72}$Ge are predicted to 
become $\gamma$-rigid and those of $^{76}$Ge as $\gamma$-soft.
For the $\gamma$-bands, the TPSM predicts moderate $\gamma$-softness in both the nuclei.
The small experimental $\gamma$-dispersion values of $^{76}$Ge indicate rigidity. 
However, this could be an artifact 
of the restricted number of matrix elements extracted from the COULEX experiment. As a matter of fact, we have
also evaluated centriods and dispersions with the same restricted set of TPSM $E2$ matrix
elements as used in the data, and obtained smaller dispersions as in the experimental work. 

It is important to point out that the comparable softness of the two isotopes derived from the $E2$ matrix elements is at variance with the staggering pattern of the
$\gamma$-bands, which is reproduced by the TPSM as well.
In the context of the collective model, the even-$I$-down pattern in $^{72}$Ge
suggest $\gamma$-softness while the even-$I$-up pattern in $^{76}$Ge
suggest $\gamma$-rigidity. The difference illustrates the limitations
of a purely collective model as compared with a microscopic approach.

\item
For the yrast- and $\gamma$-bands in the Osmium isotopes, the average triaxiality  
increases from $\langle\gamma\rangle\sim 20^\circ$ in $^{186}$Os to $\langle\gamma\rangle\sim 25^\circ$ in $^{192}$Os, with the TPSM
values 
agreeing  well with the experimental quantities. The dispersions, 
$\sigma(\textrm{cos}(3\delta))$ suggest a gradual 
increase of $\gamma$-softness along the considered isotopic chain, which
is seen in the experimental data as well. This is at variance with the 
macroscopic collective model \cite{Wilets1956,Davydov1958}, according to which the phase of the staggering parameter
associated with $^{186,188,190,192}$Os has the characteristics of being $\gamma$-soft, -rigid, -soft, -rigid,
respectively. The corresponding staggering pattern is reproduced by the TPSM calculations, while the moderate $\gamma$-softness
derived for all isotopes from the $E2$ matrix elements is consistent with the COULEX data.

\item
The TPSM predicts an average triaxiality of $\langle\gamma\rangle\sim 40^\circ$ in the yrast-band and $\langle\gamma\rangle\sim 30^\circ$
in the $\gamma$-band for $^{194}$Pt, which agrees with the experimental data.
The TPSM dispersions indicate large fluctuations in $\gamma$, while the energy 
staggering $S(I)$ shows the even-$I$-up pattern.

\item
$^{168}$Er is shown to be a prolate system with small fluctuations in $\gamma$. 

\item
The excited $0^+_2$-bands of the studied nuclei, which in the framework of the collective Bohr Hamiltonian appear as a mixture of $\gamma\gamma$ and $\beta$ 
vibrational excitations,
have been shown to be based on two-quasiparticle states. Accordingly, the transition
matrix elements to the yrast-bands and $\gamma$-bands are of single particle order 
in magnitude. 
Similarly, the structures with the band head of $I=4$ cannot be considered as collective $\gamma\gamma$ structures. These are
predominantly two-quasiparticle  states.
In both cases, the TPSM values do not correlate well with the experimental 
values of these retarded transitions. The present version of the TPSM approach seems to be
unable to describe quantitatively the fragmentation of the collective
$\gamma\gamma$-excitations among the two-quasiparticle states.

\item
For the studied nuclei, TPSM results do not reveal the correlation 
between the phase of $S(I)$ and the shape fluctuations derived from the 
$E2$ matrix elements that exists  for the collective Bohr Hamiltonian (even-$I$-down - $\gamma$-soft; even-$I$-up - $\gamma$-rigid). The
experimental dispersions,
$\sigma(\textrm{cos}(3\delta))$ are only known for the ground states of $^{76}$Ge, $^{186,188,190,192}$Os
and $^{194}$Pt, which are too uncertain to confirm the TPSM findings.

\item
The analysis of the 
individual $E2$ transition matrix elements and 
of the derived quadrupole shape invariant quantities reveals that although the   
TPSM approach generates multi-quasiparticle basis from
the Nilsson potential with a fixed triaxiality parameter $\gamma$, the subsequent configuration mixing accounts
for the  transitional shape features from 
$\gamma$-rigid to that of $\gamma$-soft. Our study adds eight examples to 
the case of $^{104}$Ru, where we demonstrated this capability of the TPSM approach for the first  time \cite{nazira}.

\end{enumerate}

We realized, after the completion of the present work, that COULEX data is also available for other nuclides which
include $^{76,80,82}$Se, $^{100}$Mo and $^{110}$Pd \cite{KAVKA1995177,PhysRevC.86.064305,D.Cline.Annu.Rev.Nucl.Sci,Hasselgren1981}. These nuclides also have sufficient
sets of $E2$ matrix elements that can be used to perform the shape invariant analysis. The TPSM results on $E2$ matrix elements and shape
invariants for these nuclides will be presented in a forthcoming publication.   

 %%%%%5%%%%%
 \section{ACKNOWLEDGEMENTS}
 One of the authors (NN) would like to acknowledge Department of Science and Technology (Govt. of India) for the
 award of INSPIRE fellowship  under sanction No. DST/INSPIRE Fellowship/[IF200508].

\appendix 
% Change the table numbering for Appendix
\renewcommand{\thetable}{A\arabic{table}}
\setcounter{table}{0} % Reset table counter

\section{Tables of the E2 matrix elements}\label{E2AP}

%=====================table1===================================================================
%\begin{table}[h!]
%\LTcapwidth=0.5\textwidth
%\caption{Phase convention adopted in present calculations
%{\bf This table was incomplete. A complete table is given in Fig. 3 of NPA 607 (1996)178, I hope you used it consistently in all the comparisons}}
%\resizebox{0.4\textwidth}{!}
  %{
%
%\begin{tabular}{|c|c|}
 % \hline
  %%=======================================================================================
  %%=======================================================================================  
%$I_i \rightarrow I_f$   &       Sign Convention\\

%  \hline
%%========================================================================================
 % $I_{yrast} \rightarrow (I-2)_{yrast}$            &$+$\\
 % $I_{\gamma} \rightarrow (I-2)_{\gamma}$            &$+$\\
 % $I_{\gamma\gamma} \rightarrow (I-2)_{\gamma\gamma}$   &$+$\\
 % $I_{0^+_2} \rightarrow (I-2)_{0^+_2}$              &$+$\\
 % $I_{\gamma} \rightarrow (I-1)_{\gamma}$            &$+$\\
 % $I_{\gamma\gamma} \rightarrow (I-1)_{\gamma\gamma}$   &$+$\\
 % $I_{\gamma} \rightarrow (I-2)_{yrast}$            &$+$\\
 % $I_{\gamma} \rightarrow I_{yrast}$                &$+$\\

%\hline
%\end{tabular}
 % }
%$  \label{sign}
%\end{table}
%=====================tableE2=two==================================================================
%{\fontsize{7}{8}\selectfont
\begin{longtable*}{p{1.8cm}p{2cm}p{1.8cm}p{1.8cm}p{1.8cm}p{1.8cm}p{1.8cm}p{1.8cm}p{1.8cm}}
%% \begin{table*}[bpt!]
%% \LTcapwidth=0.4\textwidth
\caption{Reduced TPSM  E2 diagonal  matrix elements  in $^{72,76}$Ge, $^{168}$Er,$^{186,188,190,192}$Os and $^{194}$Pt. Available  experimental matrix elements are
listed in parenthesis below the TPSM values, where the errors are quoted in square brackets.}\\
  \hline\hline
%%=======================================================================================  
$I_i \rightarrow I_f$ 	&	$^{72}$Ge	&	$^{76}$Ge	&	$^{168}$Er	&	$^{186}$Os	&	$^{188}$Os	&	$^{190}$Os	&	$^{192}$Os	&	$^{194}$Pt	\\
  \hline\hline%
  \endfirsthead

%% \hline
%% \multicolumn{2}{r}{End of the long table.}\\
%% \hline
%% \endlastfoot

%% \hline
%% \multicolumn{2}{l}{Continued from the previous page.}\\

  \hline\hline
$I_i \rightarrow I_f$ 	&	$^{72}$Ge	&	$^{76}$Ge	&	$^{168}$Er	&	$^{186}$Os	&	$^{188}$Os	&	$^{190}$Os	&	$^{192}$Os	&	$^{194}$Pt	\\
  \hline\hline
  \endhead
%  \multicolumn{2}{r}{Continued on the next page \ldots}\\ 
\hline\hline
\endfoot
%%===================================================================																	
  %\hline
  
  \multicolumn{9}{c}{$I_{yrast}  \rightarrow  I_{yrast}$}\\																	
  \cline{1-9}																	
%%===================================================================																	
																	
$2_1\rightarrow 2_1$	&	-0.156	&	-0.237	&	-3.97	&	-1.6	&	-1.861	&	-1.698	&	-1.609	&	0.683	 \\
	&	(-0.16 [{$^{+0.07}_{-0.02}$}])	&	(-0.24 [2])	&	(-3.25[{$_{-0.25}^{+0.10}$ }])	&	(-1.75 [$^{+22}_{-13}$])	&	(-1.73 [$^{+19}_{-5}$])	&	(-1.25 [$^{+22}_{-13}$])	&	(-1.21 [$^{+6}_{-17}$])	&	(0.54 [$^{+8}_{-6}$])	 \\
$4_1\rightarrow 4_1$	&	-0.159	&	-0.238	&	-3.76	&	-1.965	&	-2.211	&	-1.974	&	-1.672	&	0.514	 \\
	&	(-0.14 [{$^{+0.09}_{-0.04}$}])	&	(-0.26 [{$^{+0.01}_{-0.07}$}])	&	(-3.13[{$_{-0.09}^{+0.42}$}])	&	(-2.02 [$^{+39}_{-18}$])	&	(-2.00 [$^{+9}_{-20}$])	&	(-1.28 [$^{+27}_{-19}$])	&	(-0.73 [$^{+26}_{-6}$])	&	(1.00 [$^{+12}_{-14}$])	 \\
$6_1\rightarrow 6_1$	&	-0.171	&	-0.209	&	-4.77	&	-2.191	&	-2.339	&	-1.969	&	-1.597	&	0.121	 \\
	&	(0.20 [{$^{+0.08}_{-0.25}$}])	&	(-0.23 [{$^{+0.09}_{-0.04}$}])	&	(-5.25[{$_{-0.17}^{+0.16}$ }])	&	(-1.67 [$^{+29}_{-31}$])	&	(-1.60 [$^{+18}_{-33}$])	&	(-0.91 [$^{+24}_{-15}$])	&	(-1.16 [$^{+11}_{-26}$])	&	(0.28 [$^{+12}_{-27}$])	 \\
$8_1\rightarrow 8_1$	&	-0.416	&	-0.198	&	-4.98	&	-2.299	&	-2.296	&	-1.788	&	-1.573	&	0.389	 \\
\end{longtable*}  
%}
%%=======================================================================================

%=====================tableE2=three==================================================================
\begin{longtable*}{p{1.4cm}p{2.0cm}p{2.2cm}p{1.8cm}p{1.8cm}p{1.8cm}p{1.8cm}p{1.8cm}p{1.8cm}}
%% \begin{table*}[bpt!]
%% \LTcapwidth=0.4\textwidth
\caption{Reduced TPSM E2 inter-band matrix elements for  $I\rightarrow I $ transitions in $^{72,76}$Ge, $^{186,188,190,192}$Os and $^{194}$Pt.
Available  experimental matrix elements are
listed in parenthesis below the TPSM values, where the errors are quoted in square brackets.
The sign convention of the COULEX experiments is used, where the sign of the numerical TPSM calculation is quoted in parentheses if changed. }\\
%% \resizebox{1.00\textwidth}{!}
%%   {
%% \begin{tabular}{|c|c|c|c|c|c|c|c|c|}
  \hline\hline
%%=======================================================================================  
$I_i \rightarrow I_f$ 	&	$^{72}$Ge	&	$^{76}$Ge	&	$^{168}$Er	&	$^{186}$Os	&	$^{188}$Os	&	$^{190}$Os	&	$^{192}$Os	&	$^{194}$Pt	\\
								\hline	\hline% 
  \endfirsthead
  \hline\hline
$I_i \rightarrow I_f$ 	&	$^{72}$Ge	&	$^{76}$Ge	&	$^{168}$Er	&	$^{186}$Os	&	$^{188}$Os	&	$^{190}$Os	&	$^{192}$Os	&	$^{194}$Pt	\\
  \hline	\hline
\endhead

\hline\hline
\endfoot									
								
%%===================================================================																	
  %\hline																	
  \multicolumn{9}{c}{$I_{\gamma}\rightarrow  I_{Yrast}$ }\\																	
  \cline{1-9}																	
%%=================================================================== 																	
																	
$2_2\rightarrow 2_1$	&	0.479	&	0.667	&	0.41	&	0.999	&	0.951	&	0.909	&	1.064	&	1.351	 \\
	&	(0.65 [$^{+0.01}_{-0.02}$])	&	(0.535 [$^{+0.03}_{-0.07}$])	&	(0.47[$_{-0.01}^{+0.02}$ ])	&	(0.897 [$^{+64}_{-14}$])	&	(0.865 [11])	&	(1.065 [$^{+20}_{-37}$])	&	(1.23 [$^{+34}_{-16}$])	&	(1.517 [$^{+11}_{-18}$])	 \\
$4_2\rightarrow 4_1$	&	0.309	&	0.613	&	0.64	&	1.42	&	1.395	&	1.344	&	1.432	&	1.644	 \\
	&	(0.43 [10])	&	(0.61 [1])	&	(0.72[$_{-0.02}^{+0.02}$ ])	&	(1.22 [$^{+62}_{-55}$])	&	(1.1 [3])	&	(1.435 [$^{+43}_{-45}$])	&	(1.35 [$^{+8}_{-4}$])	&	(1.51 [$^{+6}_{-5}$])	 \\
$6_2\rightarrow 6_1$	&	0.022	&	0.621	&	0.65	&	1.686	&	1.633	&	1.541	&	1.486	&	1.029	 \\
	&	(0.18 [$^{0.25}_{-0.08}$])	&	(1.2 [$^{+0.2}_{-0.1}$])	&	(0.74[$_{-0.02}^{+0.04}$ ])	&	(1.37 [$^{+9}_{-11}$])	&	(1.46 [$^{+13}_{-25}$])	&	(1.76 [$^{+20}_{-15}$])	&	(1.49 [$^{+15}_{-6}$])	&	(1.14 [$^{+11}_{-24}$])	 \\
$8_2\rightarrow 8_1$	&	0.25	&	0.609	&	0.73	&	1.374 (-)	&	1.022	&	1.521	&	1.335	&	1.001	 \\
\end{longtable*}
%=====================tableE2=four==================================================================
\begin{longtable*}{p{1.6cm}p{2.1cm}p{1.8cm}p{1.8cm}p{1.8cm}p{1.8cm}p{1.8cm}p{1.8cm}p{1.8cm}}
%% \begin{table*}[bpt!]
%% \LTcapwidth=0.4\textwidth
\caption{Reduced TPSM E2 in-band matrix elements for transitions $I_i \rightarrow (I-2)_f$  in $^{72,76}$Ge, $^{168}$Er,$^{186,188,190,192}$Os and $^{194}$Pt. Available  experimental matrix elements are
listed in parenthesis below the TPSM values, where the errors are quoted in square brackets.
The sign convention of the COULEX experiments is used, where the sign of the numerical TPSM calculation is quoted in parentheses if changed. }\\
%% \resizebox{1.00\textwidth}{!}
%%   {
%% \begin{tabular}{|c|c|c|c|c|c|c|c|c|}
  \hline\hline
%%=======================================================================================  
$I_i \rightarrow I_f$ 	&	$^{72}$Ge	&	$^{76}$Ge	&	$^{168}$Er	&	$^{186}$Os	&	$^{188}$Os	&	$^{190}$Os	&	$^{192}$Os	&	$^{194}$Pt\\	

 \hline \hline% \midrule%
 \endfirsthead

\hline\hline%\toprule%  
$I_i \rightarrow I_f$ 	&	$^{72}$Ge	&	$^{76}$Ge	&	$^{168}$Er	&	$^{186}$Os	&	$^{188}$Os	&	$^{190}$Os	&	$^{192}$Os	&	$^{194}$Pt	\\
\hline\hline%\midrule% 
\endhead

\hline\hline%\bottomrule%
\endfoot
%%===================================================================																	
 % \hline																	
  \multicolumn{9}{c}{$I_{yrast} \rightarrow (I-2)_{yrast}$ }\\																	
  \cline{1-9}																	
%%===================================================================
%%%		Iy-(I-2)y															
$2_1\rightarrow 0_1$	&	0.402	&	0.463	&	2.33	&	1.406	&	1.6	&	1.469	&	1.491	&	1.283	\\
	&	(0.457 [4])	&	(0.526 [2])	&	(2.43[$\pm$0.07])	&	(1.674 [$^{+25}_{-21}$])	&	(1.585 ({10}))	&	(1.53 [$^{+20}_{-11}$])	&	(1.456 [$^{+8}_{-9}$])	&	(1.208 [$^{+49}_{-17}$])	\\
$4_1\rightarrow 2_1$	&	0.768	&	0.72	&	3.99	&	2.304	&	2.621	&	2.407	&	2.458	&	1.921	 \\
	&	(0.90 [2])	&	(0.795 [5])	&	(3.92[$\pm$0.12])	&	(2.761 [$^{+61}_{-70}$])	&	(2.642 ({$^{+25}_{-20}$}))	&	(2.367 [$^{+80}_{-31}$])	&	(2.115 [$^{+29}_{-28}$])	&	(1.935 [$^{+21}_{-13}$])	\\
$6_1\rightarrow 4_1$	&	0.874	&	0.897	&	5.01	&	2.992	&	3.393	&	3.133	&	3.221	&	2.619	\\
	&	(1.11 [$^{+0.04}_{-0.05}$])	&	(1.11 [$^{+0.03}_{-0.02}$])	&	(5.43[$\pm$0.16])	&	(3.89 [$^{+8}_{-5}$])	&	(3.31 ({4}))	&	(2.97 [$^{+63}_{-40}$])	&	(2.93 [$^{+77}_{-44}$])	&	(2.9 [$^{+10}_{-4}$])	 \\
$8_1\rightarrow 6_1$	&	0.037	&	0.904	&	5.63	&	3.567	&	4.074	&	3.787	&	3.853	&	3.095	\\
\end{longtable*}
%=====================tableE2=five==================================================================
\begin{longtable*}{p{1.8cm}p{1.8cm}p{1.8cm}p{1.8cm}p{1.8cm}p{1.8cm}p{1.8cm}p{1.8cm}p{1.8cm}}
%% \begin{table*}[htb!]
%% \LTcapwidth=0.4\textwidth
\caption{Reduced TPSM  E2 in-band matrix elements for transitions $I_i \rightarrow (I-1)_f$ in $^{72,76}$Ge, $^{168}$Er,$^{186,188,190,192}$Os and $^{194}$Pt. Available  experimental matrix elements are
listed in parenthesis below the TPSM values, where the errors are quoted in square brackets.
The sign convention of the COULEX experiments is used, where the sign of the numerical TPSM calculation is quoted in parentheses if changed. }\\
%% \resizebox{1.00\textwidth}{!}
%%           {
%%             \tiny
%% \begin{tabular}{|c|c|c|c|c|c|c|c|c|}
  \hline\hline
%%=======================================================================================  
$I_i \rightarrow I_f$ 	&	$^{72}$Ge	&	$^{76}$Ge	&	$^{168}$Er	&	$^{186}$Os	&	$^{188}$Os	&	$^{190}$Os	&	$^{192}$Os	&	$^{194}$Pt	\\

\hline\hline% \midrule%
 \endfirsthead

\hline\hline%\toprule%  
$I_i \rightarrow I_f$ 	&	$^{72}$Ge	&	$^{76}$Ge	&	$^{168}$Er	&	$^{186}$Os	&	$^{188}$Os	&	$^{190}$Os	&	$^{192}$Os	&	$^{194}$Pt	\\
\hline\hline%\midrule% 
\endhead

\hline\hline%\bottomrule%
\endfoot  
																	
%%===================================================================																	
%  \hline																	
  \multicolumn{9}{c}{$I_{\gamma} \rightarrow (I-1)_{\gamma}$ }\\																	
  \cline{1-9}																	
%%=================================================================== 		%%%		Ig-(I-1)g															
$3_1\rightarrow 2_2$	&	0.733	&	0.794	&	4.33	&	2.213	&	2.536	&	2.324 (-)	&	2.355 (-)	&	1.033	 \\
	&	(1.19 [2])	&	(0.52 [$^{+0.02}_{-0.04}$])	&	(3.4[$_{-0.4}^{+0.1}$] )	&		&		&		&		&		\\
$4_2\rightarrow 3_1$	&	0.165	&	0.599	&	4.45	&	2.068	&	2.383	&	2.173 (-)	&	  2.017 (-)	&	1.062	\\
	&		&	(0.64 [$^{+0.03}_{-0.07}$])	&	(5.0[$_{-0.2}^{+0.6}$])	&		&		&		&		&		 \\
$5_1\rightarrow 4_2$	&	0.397	&	0.772	&	3.26	&	2.073	&	2.386	&	2.222	&	2.245	&	1.024	\\
	&		&	(-0.9 [$^{+0.7}_{-0.2}$])	&	(4.7[$_{-0.5}^{+0.5}$])	&		&		&		&		&		 \\
$6_2\rightarrow 5_1$	&	0.021	&	0.399	&	2.75	&	1.722	&	1.925	&	1.702	&	1.441	&	0.104	\\
\end{longtable*}
%=====================tableE2=six==================================================================
\begin{longtable*}{p{1.2cm}p{1.98cm}p{2.08cm}p{2.0cm}p{1.9cm}p{1.9cm}p{1.65cm}p{2.1cm}p{2.0cm}}
%% \begin{table*}[bpt!]
%% \LTcapwidth=0.4\textwidth
\caption{Reduced TPSM E2 inter-band matrix elements for transitions $I_i \rightarrow (I-2)_f$ in $^{72,76}$Ge, $^{168}$Er,$^{186,188,190,192}$Os and $^{194}$Pt. Available  experimental matrix elements are
listed in parenthesis below the TPSM values, where the errors are quoted in square brackets.
The sign convention of the COULEX experiments is used, where the sign of the numerical TPSM calculation is quoted in parentheses if changed. }\\
%% \resizebox{1.00\textwidth}{!}
%%   {
%% \begin{tabular}{|c|c|c|c|c|c|c|c|c|}
  \hline\hline
%%=======================================================================================  
$I_i \rightarrow I_f$ 	&	$^{72}$Ge	&	$^{76}$Ge	&	$^{168}$Er	&	$^{186}$Os	&	$^{188}$Os	&	$^{190}$Os	&	$^{192}$Os	&	$^{194}$Pt	\\
								\hline\hline% \midrule%
 \endfirsthead

\hline\hline%\toprule%  
$I_i \rightarrow I_f$ 	&	$^{72}$Ge	&	$^{76}$Ge	&	$^{168}$Er	&	$^{186}$Os	&	$^{188}$Os	&	$^{190}$Os	&	$^{192}$Os	&	$^{194}$Pt	\\
\hline\hline%\midrule% 
\endhead

\hline\hline%\bottomrule%
\endfoot									
%===================================================================																	
%===================================================================																	
  %\hline																	
  \multicolumn{9}{c}{$I_{\gamma} \rightarrow (I-2)_{Yrast}$ }\\																	
  \cline{1-9}																	
%%===================================================================  																	
%%%		Ig-(I-2)y															
																	
$2_2\rightarrow 0_1$	&	0.017	&	0.198	&	0.35	&	0.716	&	0.691	&	0.662	&	0.62	&	0.016	\\
	&	(0.03 [1])	&	0.089[3])	&	(0.34 [$\pm$0.01])	&	(0.545 [$^{+13}_{-7}$])	&	(0.483 [$^{+2}_{-9}$])	&	(0.444 [$^{+9}_{-7}$])	&	(0.43 [$^{+8}_{-4}$])	&	(0.0888 [12])	\\
$4_2\rightarrow 2_1$	&	0.109	&	0.212 (-)	&	0.31	&	0.663	&	0.571	&	0.525	&	0.353	&	0.267	 \\
	&	(0.035 [6])	&	(-0.22  [$^{+0.005}_{-0.003}$])	&	(0.32 [$\pm$0.01])	&	(0.419 [$^{+27}_{-15}$])	&	(0.283 [$^{+8}_{-7}$])	&	(0.203 [7])	&	(0.13 [$^{+5}_{-8}$])	&	(0.22 [9])	\\
$6_2\rightarrow 4_1$	&	0.137	&	0.231 (-)	&	0.28	&	0.645	&	0.482	&	0.373	&	0.326	&	0.374 (-)	 \\
	&	(0.178 [$^{+0.020}_{-0.007}$])	&	(-0.186 [$^{+0.03}_{-0.005}$])	&	(0.25 [$\pm$0.007])	&	($\pm$0.325 [$^{+20}_{-26}$])	&	($\pm$0.127 [$^{+6}_{-12}$])	&	(0.195 [$^{+75}_{-74}$])	&	($\pm$0.069 [$^{+157}_{-73}$])	&	($\pm$0.224 [$^{+17}_{-19}$])	\\
$8_2\rightarrow 6_1$	&	0.181	&	0.201 (-)	&	0.22	&	0.787 (-)	&	0.426	&	0.263	&	0.654	&	0.264	\\
\end{longtable*}
%=====================tableE2=seven==================================================================
\begin{longtable*}{p{1.8cm}p{1.8cm}p{1.8cm}p{1.8cm}p{1.8cm}p{1.8cm}p{1.8cm}p{1.8cm}p{1.8cm}}

%% \begin{table*}[htb!]
%% \LTcapwidth=0.4\textwidth
\caption{Reduced TPSM E2 inter-band matrix elements for transitions $I_i \rightarrow (I-1)_f$ in $^{72,76}$Ge, $^{168}$Er,$^{186,188,190,192}$Os and $^{194}$Pt . 
Available  experimental matrix elements are
listed in parenthesis below the TPSM values, where the errors are quoted in square brackets.
The sign convention of the COULEX experiments is used, where the sign of the numerical TPSM calculation is quoted in parentheses if changed. }\\
%% \resizebox{1.00\textwidth}{!}
%%   {
%% \begin{tabular}{|c|c|c|c|c|c|c|c|c|}
  \hline\hline
%%=======================================================================================  
$I_i \rightarrow I_f$ 	&	$^{72}$Ge	&	$^{76}$Ge	&	$^{168}$Er	&	$^{186}$Os	&	$^{188}$Os	&	$^{190}$Os	&	$^{192}$Os	&	$^{194}$Pt	\\

\hline\hline% \midrule%
 \endfirsthead

\hline\hline%\toprule%  
$I_i \rightarrow I_f$ 	&	$^{72}$Ge	&	$^{76}$Ge	&	$^{168}$Er	&	$^{186}$Os	&	$^{188}$Os	&	$^{190}$Os	&	$^{192}$Os	&	$^{194}$Pt	\\
\hline\hline%\midrule% 
\endhead

\hline\hline%\bottomrule%
\endfoot  
																	
%===================================================================																	
  %\hline																	
  \multicolumn{9}{c}{$I_{\gamma} \rightarrow (I-1)_{Yrast}$ }\\																	
  \cline{1-9}																	
%%===================================================================																	
%%%		Ig-(I-1)y															
$3_1\rightarrow 2_1$	&	-0.01	&	0.349	&	-0.44	&	1.134	&	-1.095	&	1.049	&	0.979	&	0.084	\\
	&		&	(0.082 [5])	&		&		&		&		&		&		\\
$5_1\rightarrow 4_1$	&	0.087	&	0.442	&	-0.48	&	1.279	&	-1.165	&	-1.099	&	-0.894	&	-0.293	\\
	&		&	(-0.08 [$^{+0.09}_{-0.05}$])	&		&		&		&		&		&		\\
$7_1\rightarrow 6_1$	&	-0.017	&	0.469	&	-0.96	&	1.371	&	-1.175	&	-1.058	&	-0.847	&	-0.233	\\
	&		&		&		&		&		&		&		&		\\
$9_1\rightarrow 8_1$	&	-0.309	&	0.486	&	-0.91	&	1.422	&	1.153	&	0.979	&	0.887	&	0.216	 \\
\end{longtable*}
%=====================tableE2=eight==================================================================
\begin{longtable*}{p{1.8cm}p{1.8cm}p{1.8cm}p{1.8cm}p{1.8cm}p{1.8cm}p{1.8cm}p{1.8cm}p{1.8cm}}
%% \begin{table*}[htb!]
%% \LTcapwidth=0.4\textwidth
\caption{Reduced TPSM E2 inter-band matrix elements for transitions $I_i \rightarrow (I+1)_f$ in $^{72,76}$Ge, $^{168}$Er,$^{186,188,190,192}$Os and $^{194}$Pt. 
Available  experimental matrix elements are
listed in parenthesis below the TPSM values, where the errors are quoted in square brackets.
The sign convention of the COULEX experiments is used, where the sign of the numerical TPSM calculation is quoted in parentheses if changed. }\\

%% \resizebox{1.00\textwidth}{0.45\textwidth}
%% {
%% \begin{tabular}{|c|c|c|c|c|c|c|c|c|}
  \hline\hline
%%=======================================================================================  
$I_i \rightarrow I_f$ 	&	$^{72}$Ge	&	$^{76}$Ge	&	$^{168}$Er	&	$^{186}$Os	&	$^{188}$Os	&	$^{190}$Os	&	$^{192}$Os	&	$^{194}$Pt	\\

\hline\hline% \midrule%
 \endfirsthead

\hline\hline%\toprule%  
$I_i \rightarrow I_f$ 	&	$^{72}$Ge	&	$^{76}$Ge	&	$^{168}$Er	&	$^{186}$Os	&	$^{188}$Os	&	$^{190}$Os	&	$^{192}$Os	&	$^{194}$Pt	\\
\hline\hline%\midrule% 
\endhead

\hline\hline%\bottomrule%
\endfoot
																	
%===================================================================																	
  %\hline																	
  \multicolumn{9}{c}{$I_{\gamma} \rightarrow (I+1)_{Yrast}$ }\\																	
  \cline{1-9}																	
%%=================================================================== 																	
%%%		Ig-IP1y full															
$3_1\rightarrow 4_1$	&	-0.017	&	-0.218	&	-0.53	&	0.004	&	-0.024	&	-0.042	&	-0.015	&	-0.621	 \\
	&		&	(-0.44 [$^{+0.08}_{-0.05}$])	&	(-0.44 [$_{-0.09}^{+0.11}$])	&		&		&		&		&		\\
$5_1\rightarrow 6_1$	&	-0.298	&	0.275	&	-0.89	&	-1.209	&	-1.344	&	-1.352	&	-1.54	&	0.079	 \\
	&		&		&	(-0.76 [$_{-1.2}^{+0.7}$])	&		&		&		&		&		\\
$7_1\rightarrow 8_1$	&	0.309	&	-0.297	&	-0.97	&	-1.472	&	-1.664	&	-1.678	&	-0.91	&	0.254	\\
	&		&		&	(-1.2 [$_{-0.4}^{+0.9}$])	&		&		&		&		&		 \\
$9_1\rightarrow 10_1$	&	0.273	&	0.253	&	-0.95	&	-1.676	&	1.886	&	0.737	&	1.894	&	-0.005	\\
\end{longtable*}
%%%==================================================================
%=====================tableE2=nine==================================================================
\begin{longtable*}{p{1.4cm}p{1.8cm}p{2.1cm}p{2.0cm}p{1.7cm}p{1.7cm}p{1.8cm}p{2.0cm}p{2.0cm}}
%% \begin{table*}[htb!]
  % \LTcapwidth=0.4\textwidth
\caption{Reduced E2 inter-band matrix elements for transitions $I_i \rightarrow (I+2)_f$ in $^{72,76}$Ge, $^{168}$Er,$^{186,188,190,192}$Os and $^{194}$Pt isotopes. Experimental and associated errors are in parenthesis. The signs in parenthesis behind  the TPSM numbers correspond to the signs of the numerical TPSM eigenvectors. The signs in front of the numbers correspond
to the modified signs of the TPSM eigenvectors (see text) and are to be used in comparing with the experimental matrix elements. }\\
%% \resizebox{1.00\textwidth}{0.50\textwidth}
%%   {
%% \begin{tabular}{|c|c|c|c|c|c|c|c|c|}
  \hline\hline
%%=======================================================================================  
$I_i \rightarrow I_f$ 	&	$^{72}$Ge	&	$^{76}$Ge	&	$^{168}$Er	&	$^{186}$Os	&	$^{188}$Os	&	$^{190}$Os	&	$^{192}$Os	&	$^{194}$Pt	\\

 \hline\hline% \midrule%
 \endfirsthead

\hline\hline%\toprule%  
$I_i \rightarrow I_f$ 	&	$^{72}$Ge	&	$^{76}$Ge	&	$^{168}$Er	&	$^{186}$Os	&	$^{188}$Os	&	$^{190}$Os	&	$^{192}$Os	&	$^{194}$Pt	\\
\hline\hline%\midrule% 
\endhead

\hline\hline%\bottomrule%
\endfoot 
%===================================================================																	
  %\hline																	
  \multicolumn{9}{c}{$I_{\gamma} \rightarrow (I+2)_{Yrast}$ }\\																	
  \cline{1-9}																	
%%===================================================================																	
%%%		Ig-IP2y full															
$2_2\rightarrow 4_1$	&	-0.121	&	-0.062	&	0.106	&	0.004	&	-0.033	&	-0.051	&	-0.044	&	0.131	 \\
	&	(-0.06 [$^{+0.03}_{-0.04}$])	&	(0.09 (2))	&	(0.11 [$_{-0.005}^{+0.004}$])	&	([-0.5, 0.5])	&	(0.378 [$^{+50}_{-63}$])	&	(0.19 [$^{+12}_{-9}$])	&	(0.35 [$^{+16}_{-4}$])	&	(0.25 [$^{+14}_{-6}$])	\\
$4_2\rightarrow 6_1$	&	0.148	&	0.034	&	0.23	&	-1.209	&	0.239	&	0.598	&	0.331	&	0.757	 \\
	&	(0.28 [$^{+0.10}_{-0.05}$])	&	(0.35 [$^{+0.05}_{-0.03}$])	&	(0.21 [$_{-0.02}^{+0.05}$ ])	&	(0.67 [$^{+30}_{-12}$])	&	(0.57 [$^{+7}_{-12}$])	&	(0.66 [$^{+26}_{-8}$])	&	(0.40 [+9] )	&	(0.16 [$^{+6}_{-16}$])	\\
$6_2\rightarrow 8_1$	&	0.439	&	-0.053	&	0.37	&	-1.472	&	0.695	&	0.712	&	0.408	&	0.313	\\
	&		&	(-0.3 [$^{+0.2}_{-0.3}$])	&	(0.38 [$_{-0.05}^{+0.12}$ ])	&		&		&		&		&		 \\
$8_2\rightarrow 10_1$	&	0.37	&	-0.184	&	0.39	&	-1.676	&	0.743	&	0.705	&	-0.139	&	0.26	\\
\end{longtable*}
%%%=========
\bibliographystyle{apsrev4-1}
\bibliography{E2_TPSM}
%\vspace{2cm}

\end{document}